\documentclass[mnsc]{informs3}

\OneAndAHalfSpacedXI 

\usepackage{natbib}
 \bibpunct[, ]{(}{)}{,}{a}{}{,}%
 \def\bibfont{\small}%

\newcommand{\cR}{\mathcal{R}}

 \newcommand{\cQ}{\mathcal{Q}}
 \newcommand{\cS}{\mathcal{S}}
\newcommand{\cP}{\mathcal{P}}
\newcommand{\cM}{\mathcal{M}}
\newcommand{\cO}{\mathcal{O}}
\newcommand{\bean}{\begin{eqnarray*}}
\newcommand{\eean}{\end{eqnarray*}}
\newcommand{\bea}{\begin{eqnarray}}
\newcommand{\eea}{\end{eqnarray}}
\newcommand{\beaa}{\begin{array}}
\newcommand{\eeaa}{\end{array}}

\usepackage{thmtools,thm-restate}

\TheoremsNumberedThrough     
\ECRepeatTheorems

\EquationsNumberedThrough    

\begin{document}

\RUNAUTHOR{Gallego and Berbeglia}

\RUNTITLE{Bounds, Heuristics, and Prophet Inequalities}

\TITLE{The Limits of Personalization in Assortment Optimization}

\ARTICLEAUTHORS{%
\AUTHOR{Guillermo Gallego}
\AFF{School of Data Science, The Chinese University of Hong Kong, Shenzhen (CUHK-Shenzhen), Guangdong 518116, China
\EMAIL{gallegoguillermo@cuhk.edu.cn}} 
\AUTHOR{Gerardo Berbeglia}
\AFF{Melbourne Business School, The University of Melbourne, Australia \EMAIL{g.berbeglia@mbs.edu}}
} 

\ABSTRACT{%
Abstract: To study the limits of personalization, we introduce the notion of a clairvoyant firm  that can read the mind of consumers and sell them the highest revenue product that they are willing to buy. We show how to compute the expected revenue of the clairvoyant firm for a class of rational discrete choice models, and develop prophet-type inequalities that provide performance guarantees for the expected revenue of the traditional assortment optimization firm (a TAOP firm) relative to the clairvoyant firm, and therefore to any effort to personalize assortments.  In particular, we show that the expected revenue of the clairvoyant firm cannot exceed twice the expected revenue of the TAOP for the RCS model, the MNL, the GAM and the Nested-Logit Model. On the other hand, there are random utility models for which personalized assortments can earn up to $n$ times more than a TAOP firm, where $n$ is the number of products. Our numerical studies indicate that when the mean utilities of the products are heterogeneous among consumer types, and the variance of the utilities is small, firms can gain substantial benefits from personalized assortments. We support these observations, and others, with theoretical findings.  While the consumers’ surplus can potentially be larger under personalized assortments, clairvoyant firms with pricing power can extract all surplus, and earn arbitrarily more than traditional firms that optimize over prices but do not personalize them. For the price-aware MNL, however, a clairvoyant firm can earn at most $\exp(1)$ more than a traditional firm.

}


\KEYWORDS{prophet inequalities, assortment optimization, revenue-ordered assortments} \HISTORY{The first version of this paper was uploaded to ArXiv in September 2021.}

\maketitle

%




\section{Introduction}

Many e-commerce firms are moving towards offering personalized assortments based on what they know about consumers. The aim of this paper is to shed light on the magnitude of this opportunity by introducing a clairvoyant firm to upper bound the benefits of firms due to personalization. The impact of personalizing assortments on consumers is also of interest to regulators, who may wonder whether personalization is extractive.

The opportunity to personalize assortments has increased as online retailing takes away market share from traditional brick-and-mortar retailing, where personalization is more difficult to practice. Indeed, assortments offered by brick-and-mortar stores are of a more strategic nature as they are designed to show the products in an attractive way to lure consumers into the store. Changing the offered assortment requires reorganizing the store and having a backroom to hide the products that the store currently does not desire to offer. In contrast, an online platform can decide what products to display in real time based on information it gathers about consumers, such as location, search keywords, click-through paths, past purchases, and other personal attributes. Given this level of information, e-commerce firms can cluster consumers into types, and offer each type a potentially different assortment. 

Several papers addressed personalized assortment optimization policies and studied their benefits and limitations  e.g., \citep{golrezaei2014real}, \cite{bernstein2015dynamic}, \citep{el2021joint}. The proliferation of firms using personalized assortments has raised  concerns from the public and from policy makers \citep{tucker2014social}, \citep{goldfarb2012shifts} and to welfare studies of  personalized assortments and personalized pricing \citep{ichihashi2020online}.

In this paper we take a step to understand the impact of personalization when it is taken to a high degree by introducing a clairvoyant firm that manages to sell to each arriving consumer the {\em highest} revenue product that they are {\em willing} to buy. Although firms may never develop the power to read consumer minds, the analysis of this extreme case is useful as it provides a quantifiable limit to the firm's benefits of doing any personalization strategy. Moreover, the resulting upper bounds are elegant, tight, and, for some families of choice models, only a constant factor away from the expected revenue of a traditional assortment optimization (TAOP) firm.\footnote{By the traditional optimization problem we mean the {\em unconstrained} assortment optimization problem, where a {\em single} assortment of arbitrary size is offered to all consumer types.} We find that the consumer's surplus may actually be larger under a clairvoyant firm than under a TAOP firm. This is due, in part, to the willingness of the clairvoyant firm to sell products not offered by the TAOP firm.  Our finding suggests that regulators need to be extremely careful in regulating personalized assortments.

\subsection{Our Contributions}


\begin{itemize}

\item \textbf{Clairvoyant firm in assortment optimization}:
To explore the limits of personalization, we introduce a clairvoyant firm that sells to each arriving consumer the highest revenue product she is willing to buy. We demonstrate that the expected revenue of a clairvoyant firm serves as an upper bound on the expected revenue achievable by any personalized policy. In Theorem~\ref{thm:fcr} we show that under mild conditions, the expected revenue of a clairvoyant firm can be computed in closed form with the mere knowledge of the outside alternative probabilities for revenue-ordered assortments. Furthermore, in Theorem~\ref{thm:newups} we establish upper bounds on the revenue achievable by the clairvoyant firm, relying solely on the vector of last-choice probabilities \footnote{Under a discrete choice model $\mathcal{P}$, the last choice probability for an alternative $i$ is $\mathcal{P}(i,\{i\})$ (i.e. the probability of choosing $i$ when $i$ is the only alternative being offered.)}.

We have also examined extreme cases and discovered intriguing results. First, we demonstrate that for regular Discrete Choice Models (DCMs), a clairvoyant firm has the potential to earn up to $n$ times more revenue than a TAOP firm, where $n$ represents the number of products available. Remarkably, this bound is tight and there is a personalized policy that earns as much as the clairvoyant firm. Moreover, the result remains true even when considering the restricted setting of the Markov chain choice model (Theorem~\ref{thm:ubn}). At the other end of the spectrum, when the DCM is governed by the {\em independent} choice model, a clairvoyant firm can make no more than a traditional firm that does not personalize assortments (Proposition~\ref{prop:idm}). 


\item \textbf{Prophet Inequalities for Assortment Optimization}:
We introduce the well-known prophet inequality into the field of assortment optimization. We first investigate DCMs with independent value-gaps, and demonstrate the validity of the prophet inequality and prove that a clairvoyant firm that can personalize assortments can generate at most twice the revenue of a TAOP firm (Proposition~\ref{prop:pi}). Models with independent value-gaps include the Random Consideration Set (RCS) model, and the random utility model with deterministic utility for the outside alternative. We also present sufficient conditions for the prophet type inequalities to hold for models with dependent value gaps, showcasing that the prophet inequality holds for the Multinomial Logit (MNL) (Section \ref{sec:MNL}), the Generalized Attraction Model (GAM) (Section \ref{sec:GAM}), and the Nested Logit (NL) model\footnote{The results for the NL model are relative to a nest-clairvoyant firm.} (Section \ref{sec:NL}). We also studied the limits of personalization under the well-known LC-MNL model and the Markov Chain model. For the LC-MNL model, we show that a clairvoyant firm cannot earn no more than twice as much as a firm that can personalize assortments for each segment. For the Markov chain model, we prove a rather strong negative result: a firm that can customize assortments based on the consumer's first preference does not yield any additional benefits. In contrast, in Section \ref{sec:examples} we highlight models where personalizing based on the consumer's first choice yields significant benefits. These findings highlight the implications of the prophet inequality and personalization in different model frameworks.

\item \textbf{Experiments}:
We conducted a numerical study that examines the bounds of personalization in randomly generated latent-class MNL models. These results suggest that the most advantageous scenarios for personalization are those where the firm possesses knowledge of the underlying MNL model for each consumer type, the coefficient of variation of the utilities is small, and there is sufficient spread in the mean utilities across consumer types. This is consistent with our intuition as then knowing the type carries significant information. On the other hand, for instances with low heterogeneity in the means and high variance in the utility, the benefits of personalization become small or
negligible. These insights are confirmed by analytical observations and Theorem~\ref{thm:bigbeta}.

\item \textbf{Extensions}:
We investigate clairvoyant pricing and explore its implications. Specifically, we demonstrate that a clairvoyant firm with the ability to customize its prices to each customer has the potential to earn arbitrarily more revenue than a TAOP firm (Proposition~\ref{prop:pricing_unbounded}), highlighting the significant advantage of perfect information coupled with pricing freedom. However, for the Multinomial Logit (MNL) model, we establish that a clairvoyant firm that can customize the prices to each incoming consumer can make up to $e \approx 2.71828$  times more revenue with respect to a firm that must treat all consumers equally (Proposition~\ref{prop:MNLprice}).
Additionally, we provide personalization results that relate to recent papers on refined assortment optimization \citep{berbeglia2021refined} and joint assortment and customization \citep{el2021joint}. These new theorems offer refinements to those works with new bounds provided by the clairvoyant firm. 
\end{itemize}

\subsection{Related literature}

The literature on assortment optimization has increased dramatically during the last 15 years starting with the seminal paper of \citet{talluri2004revenue} where the authors assume that consumer preferences can be described by an MNL model. Reviews of the subject can be found in \citet{strauss2018review}, \citet{den2015dynamic}, \citet{lobel2021revenue}, and the book by \cite{gallego2019revenue}. The assortment optimization problem has been studied under different choice models (see, e.g., \citet{blanchet2016markov} and \citet{davis2014assortment}). In addition, it has also been studied in different settings such as where the firm faces cardinality limitations on the offer sets and similar constraints (see e.g., \citet{rusmevichientong2010dynamic,desir2020constrained,sumida2021revenue}) and in settings where inventory is limited (see e.g., \citet{topaloglu2009using}). While most of the literature focuses on a single firm, some papers have studied assortment equilibria when there are two or more competing firms \citep{besbes2016product, nip2022competitive,  aouad2021algorithmic}. Recently, there has been an interest in understanding the limitations of traditional assortment optimization and assessing the benefits of enlarging the possible actions taken by the firm such as using lotteries \citep{ma2021assortment}, reducing product utilities \citep{berbeglia2021refined}, increasing product advertisement \citep{wang2021advertising}, and personalizing the assortment offered (see e.g., \cite{bernstein2015dynamic} and references in the following paragraph).

While the benefits of personalization have been recognized several decades ago (see, e.g., \citet{surprenant1987predictability}), it is only recently that researchers began to study personalized assortment optimization problems. In these problems, consumers are divided into types, and each type follows a discrete choice model that has residual uncertainty. The objective is to choose a (possibly) different assortment to offer to each segment to maximize expected revenues. One of the earliest works in personalized assortments was carried out by \citet{bernstein2015dynamic} who studied a finite-horizon setting in which consumers follow a mixed MNL model, demand is stationary, and the seller is able to observe the segment class of the incoming consumer. In their model, all products have the same revenue, products are limited in inventory, and the seller must choose a personalized assortment at each period. The authors provide structural results about the optimal policy and develop some heuristics. \citet{chan2009stochastic} study a framework that contains the previous model but allows non-stationary demand and different product prices and show that a myopic policy guarantees at least half of the revenue of the optimal control policy. \citet{golrezaei2014real} also extended the model of \citet{bernstein2015dynamic} to non-stationary demand allowing different prices and proposed a personalized assortment policy that achieves 50 percent of the optimal revenue even against an adversarial chosen demand. \citet{li2023deep} developed a deep reinforcement learning algorithm for the same problem and they showed experimentally that it performs better than the other existing methods. \citet{gallego2015online} consider a similar problem but allow for product revenues to be dependent on the consumer types. The authors propose online algorithms to offer personalized assortments that guarantee a factor of $(1-\epsilon)$ of the optimal offline  revenue (under complete information) where $\epsilon$ is the error in computing an optimal solution to the choice based linear program (CDLP).
\cite{bernstein2019dynamic} propose an exploration-exploitation framework to learn consumer preferences and personalize assortments under a finite-horizon. They develop a dynamic clustering estimation algorithm that maps consumer segments to clusters. In a case study, the authors show that the clustering policy increased transactions by more than 37\% with respect to learning and treating each consumer type separately. \cite{kallus2020dynamic}, who consider a similar framework, argue that the amount of data required to estimate a LC-MNL model is orders of magnitude larger than the data available in practice. To overcome this issue, they imposed that the parameter matrix associated with the LC-MNL has a low rank and show that the model can be learned quickly. They have also shown that an exploration-exploitation algorithm that is rank aware and does assortment personalization has much lower regret with respect to those who ignore the rank structure. \cite{cheung2017thompson} study another  exploration-exploitation setting in which each consumer follows its own MNL model according to their observable attributes.  They developed a Thompson sampling-based policy to personalize assortments and prove regret bounds with respect to the optimal policy. \citet{jagabathula2020personalized} develop algorithms to perform personalized promotions in real time. The authors consider a choice model in which consumers have a partial order among the products that are combined with an MNL. They develop a MILP which, for an incoming consumer, would personalize the assortment of products offered at a fixed discounted price. \cite{chen2021statistical} consider a learning problem where a firm uses transactions to personalize prices or assortments. The authors developed a unified logit modeling framework in which products and consumers have a feature vector that lies in a multi-dimensional real space. The nominal value of a product to a given consumer is a linear function of the product and consumer features and the error terms follow a Gumbel distribution. They establish finite-sample convergence guarantees that are later traduced into out-of-sample performance bounds. Recently, \citet{chen2023assortment} studied how firms should personalized product recommendations once a customer has decided to purchase a product. The authors developed a novel model that can account for inventory constraints and provide a constant factor approximation algorithm. \citet{ettl2020data} studied the problem of offering customized bundles and  developed a data-driven algorithm that accounts for inventory restrictions. 

A personalized assortment may reveal private customer data about the consumer to third-parties. Recently, \citet{lei2020privacy} consider the personalized assortment optimization problem when the firm must ensure that the assortment policy doesn't reveal private information using the differential privacy framework \citep{dwork2006differential}. \citet{berbeglia2021refined} provide tight revenue guarantees on the performance of the well-known revenue-ordered assortment strategy with respect to the optimal personalized assortment solution. Their result holds for regular DCMs (which includes all RUMs) and works even under \emph{personalized refined assortment optimization} where the firm may reduce the product utilities to some consumer types. \cite{el2021joint} studies a two-stage personalized assortment optimization problem with capacity constraint under the LC-MNL model. In their model, consumers follow an LC-MNL and the firm is able to observe the segment of the incoming consumer to customize the final assortment offered. After proving that the problem is NP-hard, they developed an efficient algorithm that guarantees $\Omega(\frac{1}{\log(m)})$-fraction of the optimal revenue where $m$ is the number of segments. We strengthen that result and show that the same revenue guarantee holds with respect to a clairvoyant firm (see Section \ref{sec:joint_assortment}). More recently, \citet{udwani2021submodular} provided a $(0.5-\epsilon)$-approximation algorithm for the same problem. Recently, \cite{goyal2023mnl} consider a different stochastic variant of the assortment optimization problem, where the parameters that determine the revenue and the demand of each item are drawn from some known distribution. While the authors also evaluate the profit of a clairvoyant firm, their setting is very different from ours since the randomness in their case comes from the model parameters (including revenue) and not from the consumer choices \footnote{While their setting is quite different, the techniques used in their paper are similar to those proposed in an earlier version of this paper \citep{gallegoberbegliaarxiv2021}.}.

Many researchers have studied settings where the firm can customize product prices.  One key advantage of a personalized assortment policy with respect to personalized pricing is that it is easier to implement as there is no need to calibrate a price-aware discrete choice model. In addition, personalized pricing is sometimes banned by law\footnote{For example, Tinder settled a class action lawsuit for \$17.3 million for charging higher prices to people over 30 years old. URL: \url{https://www.theverge.com/2019/1/25/18197575/tinder-plus-age-discrimination-lawsuit-settlement-super-likes}} and it is generally perceived as an unfair practice \citep{haws2006dynamic}. A personalized assortment strategy can better deal with those issues. For instance, a firm doing personalized assortments may simply personalize the products that appear at a prominent position (e.g., on the first page of results) but allow consumers to see the same set of products if they continue browsing. Although the offer set is actually the same for all consumers, this policy has a similar effect in consumers as personalized assortments \citep{abeliuk2016assortment,gallego2020approximation,aouad2021display,berbeglia2021market,derakhshan2022product}. The reader interested in personalized pricing is referred to \citet{elmachtoub2021value}, \citet{chen2020privacy}, \citet{gallego2021bounds}, and \citet{fallah2024limits} and references therein. In Section \ref{sec:pricing} we prove that; under the MNL model; a clairvoyant firm who can customize product prices for each consumer can extract up to $e\approx 2.718$ times more revenue with respect to the firm who must set the same prices to each consumer.

\section{Assortment Optimization, Personalization, and Clairvoyant Firms}

In this section we will briefly review discrete choice models (DCMs) and the traditional assortment optimization problem (TAOP),  present some preliminary results, discuss the personalized assortment optimization problem (p-TAOP), and introduce a clairvoyant firm that can read the minds of consumers and sell them the highest revenue product they are willing to buy. 

\subsection{Review of Discrete Choice Models}
 Let $N = [n] := \{1, \cdots, n\}$ denote a set of products that the firm can offer to consumers, and let $0$ denote the outside alternative. We assume that the outside alternative is always available, so for any assortment $S \subset N$, the choice set for consumers is $S_+ : = S \cup \{0\}$. Let $\cS$ be the set of all subsets of $N$. A mapping $\cP(\cdot,\cdot): N_+ \times \cS  \rightarrow [0,1]$ is a DCM if and only if $\sum_{i \in N_+} \cP(i,S) = 1~\forall~S \subset N$, and $\cP(i,S) = 0~\forall~i~\in~N\setminus S$. For any DCM,  $\cP(i,S)$ represents the probability of selecting $i \in N_+$ from assortment $S \subset N$. A DCM is regular if $\cP(i,S)$ is weakly-decreasing in $S \ni i$.

A  DCM is said to be a random utility model (RUM) if there are continuous random variables $U_i, i \in N_+$ such that $\cP(i,S) = \\mbox{Pr}(U_i \geq U_j ~\forall j \in S_+), ~i \in S_+,~ S \subset N$. All RUMs are known to be regular \citep{mcfadden1990stochastic}. The class of RUMs is equivalent to the class of DCMs characterized by  distributions over preference orderings \citep{block1960random}.


The multinomial logit (MNL) model is a RUM that will play a central role in this paper as an auxiliary model for more general DCMs. The random utilities of an MNL model are assumed to be independent Gumbel random variables with location parameters $u_i, i \in N_+$, and a common scale parameter $\beta$.  The location parameter of the outside alternative is normalized to zero. Let $v_i := \exp(u_i/\beta), i \in N_+$. Let $v := (v_i)_{i \in N}$ be the vector of the products' attraction values with $v_0 = \exp(0) = 1$. We will use the notation $\cM_v(i,S) := v_i/(1 + \sum_{j \in S}v_j), i \in S_+, S \subset N$ to denote the choice probabilities under the MNL with attraction vector $v$. The MNL is an instance of the basic attraction model (BAM) of Luce \citep{luce1958probabilistic} that emerges also from the representative agent problem with entropy concentration costs. 


\subsection{Review of the Traditional Assortment Optimization Problem}
Let $r_i$ be the revenue\footnote{$r_i$ can also be the profit contribution defined as the spread between the unit price and the unit cost of the product.} associated with the sale of one unit of product $i \in N$. We assume without loss of generality that the products are labeled in decreasing order of their revenues, so $r_1  \geq \cdots \geq r_n > 0$. For convenience we also define $r_0:= 0$. For any DCM $\cP$, the expected revenue associated with assortment $S \subset N$ is given by $R(S) := \sum_{i \in S}r_i \cP(i, S)$. The problem of finding an assortment $S \subset N$ that maximizes $R(S)$ is known as the traditional assortment optimization problem (TAOP).  We call a firm that faces the TAOP, a TAOP-firm. We will denote the optimal expected revenue by
$\cR^* := \max_{S \subset N}R(S)$, and an optimal assortment by $S^* \in \arg\max_S R(S)$. The TAOP is NP-hard\footnote{In fact, for general RUMs it is NP-hard to approximate TAOP to within a factor of $\Omega(1/n^{1-\epsilon})$ for every $\epsilon>0$ \citep{aouad2018approximability}.} although polynomial algorithms exist for some models. For the MNL, the TAOP has a well-known solution summarized in the following proposition. All omitted proofs can be found in the Appendix.

\begin{proposition}(Folklore)
\label{prop:mnl}
    For an MNL model with attraction vector $v$, the optimal expected revenue, $\cR^*_v$, is the unique root of the equation $\tau = \sum_{i \in N}(r_i - \tau)^+ v_i$.\footnote{For any real number $x$, the quantity $x^+ := \max(x,0)$.} Moreover, $\cR^*_v$ is weakly increasing in $v$, and $S^*_v = \{i \in N: r_i > \cR^*_v\}$ is an optimal assortment. 
\end{proposition}

A reasonable heuristic to use for the TAOP in choice models where it is NP-hard to find the optimal value is to limit the optimization to the class of revenue-ordered assortments $S(\tau): = \{i \in N: r_i > \tau\}$. We denote by $\cR^o := \max_{\tau > 0}R(S(\tau)) = \max_{i \in [n]}R([i])$. Performance guarantees for the revenue-ordered assortment, relative to $\cR^*$, can be found in \citet{rusmevichientong2014assortment}, \citet{aouad2018approximability}, \citet{berbeglia2020assortment} and \citet{berbeglia2021refined}.

For the rest of this section, we will consider regular choice models $\cP$ with the property that $\cP(0,N) > 0$.  This assumption fits situations where there is a positive probability that consumers will walk away without making a purchase. Indeed, by regularity $\cP(0,S) \geq \cP(0,N) > 0$ for all $S \subset N$. Let $\cO(i,S): = \cP(i,S)/\cP(0,S)~\forall~i \in S,~\forall~S \subset N$ be the odds ratio for product $i$, when assortment $S$ is offered.  The following result links the odds ratio to the assortment optimization problem, and will be helpful in establishing other results later in the paper.

\begin{lemma}
\label{lem:formulaoddsratio}
For any regular choice model with $\cP(0,N) > 0$ and for any assortment $S \subset N$,
$R(S)  = \sum_{i \in S} \cO(i,S)(r_i - R(S))$, and in particular
$\cR^* = \sum_{i \in S^*} \cO(i,S^*)(r_i - \cR^*)$.
\end{lemma}



\section{Clairvoyant firm, Heuristics, and Prophet Inequalities.}

We assume that there is a sequence of arriving consumers whose collective behavior is consistent with $\cP$. In particular, the underlying model may be of the form 
\begin{equation}
    \label{eq:pm}
\cP(i,S) = \sum_{j \in M}\theta_j \cP_j(i,S),
\end{equation}
where $M$ represents the set of consumer types,\footnote{For ease of exposition we assume a discrete distribution of types, but all results are valid for continuous types.} $\cP_j$ represents the DCM for type $j \in M$, and $\theta_j > 0, j \in M, \sum_{j \in M}\theta_j = 1$ is the distribution over types. Let $R_j(S) := \sum_{i \in S}r_i \cP_j(i,S)~\forall S \subset N, \forall j \in M$, and $\cR^*_j := \max_{S \subset N}R_j(S)$ for all $j \in M$. A firm that can identify the types of arriving consumers can personalize assortments and earn $\cR^p := \sum_{j \in M} \theta_j \cR^*_j \geq \cR^*$.

Since the underlying DCMs $\cP_j, j \in M$ may have residual uncertainties, even a firm that can personalize assortments cannot make as much revenue as clairvoyant firm that can read the minds of consumers and sell them the highest revenue product that they are willing to buy. To be more precise, we say that an arriving consumer is willing to buy product $i$ if there is an assortment $S \ni i$ such that the consumer selects $i$ when offered assortment $S$. For any $i \in S_+, S \subset N$, let $\hat{B}_i(S)$ be a Bernoulli random variable that takes value one if $i$ is chosen from $S$, and takes value zero otherwise. 

We define $B_i: = \max_{S \subset N}\hat{B}_i(S)$, and $q_i := \mbox{Pr}(B_i = 1), i \in N$. We say that the vector $q = (q_i)_{i \in N}$ is the vector of {\em maximal} probabilities. For RUMs, since $\cP(i,S) \leq \cP(i,\{i\})$, we see that $q_i = \cP(i,\{i\}) := \omega_i$, so the vector of maximal probabilities is the vector of last-choice probabilities. In general, however, $q_i$ may be strictly larger than $\omega_i$ as in the non-standard nested logit (NL) model with dissimilarity parameters greater than one.

Let $X_i: = r_i B_i$, and consider a clairvoyant firm that can observe $B_i, i \in N$, and earn $\bar{\cR} = E[\max_{i \in N} X_i]$.  We next turn to prophet inequalities of the form 
\begin{equation}
\label{eq:piphi}
 \cR^* \geq \frac{\phi}{1 + \phi}\bar{\cR}, 
\end{equation}
for $\phi > 0$, which imply that $\cR^* \geq \frac{\phi}{1 + \phi} \cR^p$.  Often we will find prophet inequalities of the form $\cR^o \geq \frac{\phi}{1 + \phi} \bar{\cR}$ which provide performance guarantees for the revenue-ordered heuristic relative to $\bar{\cR}$, and therefore potentially much stronger than  guarantees relative to $\cR^*$.

 Our first result is for the special case of independent $X_i, i \in N$, or equivalently independent $B_i, i \in N$.  For the independent case there is a well-known threshold policy for the prophet problem, \citep{lucier2017economic} that leads to (\ref{eq:piphi}) with $\phi = 1$, see \citep{krengel1977semiamarts}. It  is easy to see that threshold policies for the prophet problem translate into revenue-ordered heuristic for assortment optimization. We summarize this result in the following proposition, with a detailed proof in the Appendix.

\begin{proposition}
\label{prop:pi}
For DCM with independent $B_i, i \in N$, (\ref{eq:piphi}) holds for $\phi = 1$. Moreover, the bound is tight.
\end{proposition}

In Section \ref{sec:application} we show that some choice models satisfy the independence of $B_i$'s. However, many DCMs do not satisfy this condition. For example, the $B_i, i \in N$ are positively correlated for the MNL model. To go beyond the independent case we start by the important and well-known observation that for any random variables $X_i, i \in N$, $\max_{i \in N}X_i \leq \tau + \sum_{i \in N}(X_i - \tau)^+$ for all $\tau$. Noticing that for $X_i = r_iB_i$, we have $(X_i - \tau)^+ = (r_i - \tau)^+B_i$, we see that 
$\bar{\cR} = E[\max_{i \in N}X_i] \leq \tau + \sum_{i \in N}(r_i-\tau)^+q_i$. Evidently,  $\bar{\cR} \leq \hat{\cR} :=  \min_{\tau \geq 0} [\tau + \sum_{i \in N}(r_i-\tau)^+q_i]$.

Since $\cR^o \leq \cR^* \leq  \cR^p \leq \hat{\cR}$ we see that an upper-bound on the benefit of personalization can be obtained numerically by computing $\cR^o$ and $\hat{\cR}$ and taking the ration $\hat{\cR}/\cR^o$. This can be done in $O(n)$ time, even if computing $\cR^*$ or $\cR^p$ is NP-hard. 

To obtain further bounds we define $\tau_{\phi}$  as the unique root of the equation $\tau = \sum_{i \in N}(r_i - \tau)^+ \phi q_i$ for all $\phi > 0$.

\begin{theorem}
\label{thm:newups}
 $$\bar{\cR} \leq \frac{1 + \phi}{\phi}\tau_{\phi}~~~~\forall~~~~\phi  > 0.$$
 \end{theorem}

From these upper-bounds we can find obvious sufficient conditions for prophet inequalities. As an example, the prophet inequality (\ref{eq:piphi}) holds for $\phi = 1$ if 
$\hat{\cR} \leq 2\cR^*$ or if $\tau_1 \leq \cR^*$. Other sufficient conditions include $\hat{\cR} \leq 2\cR^o$, and  $\tau_1 \leq \cR^o$. The last two conditions, which can be verified in $O(n)$ time.

The following sufficient condition provides the key to prophet-type inequalities for many commonly used discrete choice models. 
\begin{theorem}
\label{thm:sc}
For any $\phi > 0$,  consider the revenue-ordered assortment $S_{\phi} := \{i \in N: r_i > \tau_{\phi}\}$. 
If 
\begin{equation}
\label{eq:condphi}
\cP(i,S_{\phi}) \geq \phi q_i \cP(0, S_{\phi})  ~~~~\forall~i \in S_{\phi}
\end{equation}
holds, then (\ref{eq:piphi}) holds for $\phi$. Let  $\phi^*$ be the largest $\phi > 0$ for which (\ref{eq:condphi}) holds. Then
 $$\cR^* \geq \cR^o \geq \frac{\phi^*}{1 + \phi^*} \bar{\cR}.$$
\end{theorem}

It is clear that when condition (\ref{eq:condphi}) holds for $\phi = 1$, then the classic version of the prophet inequality holds, providing a 50\% guarantee for both $\cR^*$ and $\cR^o$ relative to $\bar{\cR}$.  Verifying condition (\ref{eq:condphi}) for any $\phi$ requires $O(n)$ work assuming that the vector $p$ is known.  Consider any DCM for which the odd-ratios satisfy  the condition $\cO(i,S) \geq \cO(i, \{i\})$. Then, since $\cO(i, \{i\}) = \omega_i/(1-\omega_i) \geq \omega_i = \cP(i, \{i\})$, we see that condition (\ref{eq:condphi}) holds for $\phi = 1/(1- \min_{i \in N}w_i) \geq 1$ for models where $q = \omega$.

So far we have obtained prophet inequalities based on {\em upper} bounds on $\bar{\cR}$ bypassing the need to compute $\bar{\cR}$. However, having the ability to compute $\bar{\cR}$ provides us with a more precise assessment of the potential benefits of personalization, and a direct mechanism to compute the ratio $\cR^*/\bar{\cR}$ and $\cR^o/\bar{\cR}$. We will next present a class of DCMs that allows for an efficient computation of $\bar{\cR}$.  We say  that a DCM is {\em rational} if it is regular {\em and} satisfies assumptions R1 and R2 below. For economy of notation we say that rational DCMs satisfy {\bf Assumption R}. We remind the reader that $B_i := \max_{S \subset N}\hat{B}_i(S)$ for all $i \in N_+$. \\
\textbf{R0:} (regularity) $\cP(i,S)$ is weakly decreasing in $S \ni i$ for all $i$. \\
\textbf{R1:} (Last-choice rationality) $B_i =\hat{B}_i(\{i\})$for all $i \in N$. \\
\textbf{R2:} (No-purchase rationality) If the set $\{i \in S: B_i = 1\}$ is non-empty, then $\hat{B}_0(S) = 0$.

Last-choice rationality implies that if a product is selected from any set $S\subseteq N$, it will also be selected when it is offered alone so it is that last available (only) choice (besides the no-purchase option). No-purchase rationality implies that if there is at least one product in $S$ that the consumer is willing to buy (when offered some assortment $T$), then the consumer would not select the outside alternative when offered assortment $S$. R1 and R2 were proposed by \cite{gallegoirvani},  who show that the class of choice models that satisfy Assumption R is closed under convex combinations and includes {\em all} RUMs. Assumption R removes regular models, say $\cP$, that are decomposed into a convex combination, as in equation (\ref{eq:pm}), where some of the $\cP_j, j \in M$ fail to satisfy R1 or R2. It is clear that for rational decision models that $q_i = \omega_i = \cP(i, \{i\})$. We are now ready to show how to compute $\bar{\cR}$ under Assumption R.


\begin{theorem}
\label{thm:fcr} For any DCM satisfying R1 and R2, and therefore for any  rational DCM
\bea
\bar{\cR} = \sum_{i \in N} r_i[\cP(0,[i-1])- \cP(0,[i])]. \label{eq:fcr}
\eea
\end{theorem}

The reader can verify by simple algebra $\bar{\cR} = \sum_{i \in N}(r_i - r_{i+1})\cP([i], [i])$, by setting $r_{n+1}: = 0$, and $\cP(S,S) := 1 - \cP(0,S)$ for all $S \subset N$. Formula (\ref{eq:fcr}) induces a discrete choice model, say $\cQ$, such that for any assortment $S$, the expected revenue of the clairvoyant firm restricted to $S \subset N$ is given by $\bar{R}(S) = \sum_{i \in S}r_iQ(i,S)$. The choice model $\cQ$ is given by $\cQ(i,S) := \cP(0,[i-1]\cap S)- \cP(0,[i]\cap S),~i \in S, ~S \subset N$, with $\cQ(0,S) = \cP(0,S)$ The reader may wonder whether there is a choice model $\cP$ such that $\cQ = \cP$; the answer is yes as this is true for the independent demand model as we will verify in the next section when we deal with applications to specific choice models.

The ability to compute $\bar{\cR}$ can give significant insights to firms contemplating the use of personalized assortments. As an illustration, suppose that an online retailer has an enormous amount of information from its customers but currently does not personalize assortments. Based on all transactions, the firm has estimated a complex choice model $\cP$ that consisted of a mixture of three consumer segments, each of which has a rational DCM \footnote{For instance, a mixture choice model composed of a feature base LC-MNL model, a Markov chain choice model without features and a feature-based neural network based on the RUM model.}. Since each of the three segments satisfies rationality (Assumption R), the firm used the formula in Theorem \ref{thm:fcr} and found that $\bar{\cR}=\$ 11.61$. The firm is currently using a heuristic to solve the TAOP that has expected revenue of $\$11.09 \leq \cR^*$.  Based on this findings, the firm decides not to pursue any personalization strategy as in the best case they can increase the revenue by at most 4.7\%.\footnote{Note that the firm also learnt that the heuristic is very close to the optimal TAOP revenue and may abandon any efforts to improve it.} In Section 5, we have conducted several random experiments to see how much more the expected revenue $\bar{\cR}$ of the clairvoyant firm is relative to $\cR^*$ and $\cR^o$. We also elaborate on the factors that can lead to larger or smaller gains.

We next ask what is  theoretically the most a clairvoyant firm earn relative to a TAOP firm. It is possible to construct examples where a clairvoyant firm earns {\em arbitrarily} more than a TAOP firm over the class of all DCMs.  We next show that for rational DCMs it is possible to construct examples where the clairvoyant firm, and a firm that can personalize assortments, can both earn up to $n$ times more than the TAOP firm. We remind the reader that in the following theorem the notation $\cR^p = \sum_{j \in M} \theta_j \cR^*_j$ corresponds to the expected revenue from personalizing assortments. 

\begin{theorem}
\label{thm:ubn}
For any DCM satisfying R1 and R2, and therefore for any rational model $\cR^p \leq \bar{\cR} \leq n \cR^o \leq n \cR^*$. Moreover, for every $n$, and every $\epsilon >0$ we can construct an DCM instance with $n$ products where $\bar{\cR} = \cR^p \geq n \cR^* - \epsilon$ for all $\epsilon >0$. These results remain true even within the class of Markov chain choice models.
\end{theorem}

While Theorem~\ref{thm:ubn} shows there are examples where the firm surplus increases by a factor of $n$, this doesn't necessarily mean that consumer surplus would drop substantially. In fact, it is possible to construct examples where consumers are better off under a clairvoyant firm. This can happen because consumers who select the no-purchase alternative when offered an optimal TAOP assortment, may still buy and earn positive surplus under the clairvoyant firm.  On the other hand, consumers who originally purchased under a TAOP firm may obtain a lower surplus under a clairvoyant firm. The balance for the consumer can go either way. However, in all the worst-case examples developed in this paper, including the worst case developed in Theorem~\ref{thm:ubn}, consumer surplus increased under a clairvoyant with respect to a TAOP firm. Although more research is needed about the impact of personalized assortments on consumers' surplus, this is beyond the scope of this paper. For now we can say that although intuitive, it is not correct to say that the clairvoyant policy will lead to a lower expected surplus for consumers relative to a TAOP firm.

In what follows we will apply our results to establish prophet inequalities for a variety of models, and remark that for all of the models for which the prophet inequality holds, we also have performance guarantees for revenue-ordered heuristics some of which are new in the literature in the sense that the guarantee is relative to an upper-bound on the expected revenue of a clairvoyant firm.

\section{Application to Specific Choice Models}\label{sec:application}

We now apply the theoretical results from the previous sections to a variety of choice models that are popular in the literature.

\subsection{The Independent Demand Model.}
Under the independent demand (ID) model the choice probabilities are given by  $\cP(i, S) = \cP(i,N) = \omega_i, ~\forall~ S \ni i, S \subset N$. It is easy to see that $\cR^* = R(N) = \sum_{i \in N}r_i \omega_i$. On the other hand, $\cP(0,[i-1])- \cP(0,[i])] = \omega_i, i \in N$, and therefore, by Theorem~\ref{thm:fcr}, 
$\bar{\cR} = \sum_{i \in N} r_i\omega_i = \cR^*$, leading to the next result. 

\begin{proposition}
    \label{prop:idm}
    For the ID model $\cR^* = \bar{\cR}$.
\end{proposition}

Consequently, there is no possible benefit from personalizing assortments for the ID model. We remark that the ID model was the DCM used in early revenue management applications, and the main issue was that of designing differentiated fares, with the optimal assortment being the set of all fares exceeding the marginal value of capacity.

\subsection{DCM Models with Independent $B_i$s.}

For models with independent $B_i$s we have $\cP(0,[i]) = \Pi_{j \leq i}\cP(B_j = 0) = \Pi_{j \leq i}(1- \omega_j)$, so 
\begin{equation}
 \label{eq:rcs}
 \bar{\cR} = \sum_{i \in N}r_i[\Pi_{j \leq i-1}(1- \omega_j) - \Pi_{j \leq i}(1- \omega_j)] =  \sum_{i \in N}r_i \Pi_{j < i}(1- \omega_j) \omega_i.
 \end{equation}

By Proposition~\ref{prop:pi} we have 
$$\bar{\cR} \leq 2 \cR^o \leq 2 \cR^*$$
for all models with independent $B_i, i \in N$.

This result applies, for example, to any  RUM with independent $U_i, i \in N$, and deterministic $U_0$. The special case where $U_i = u_i + \epsilon_i$,  and the $\epsilon_i, i \in N$ are independent Gumbel random variables was analyzed by \cite{wang2021impact}. The author has shown that the assortment optimization problem is NP-hard. Our result guarantees that the revenue-ordered heuristic, $\cR^o$, achieves at least $0.5 \bar{\cR}$ regardless of the distribution of the $\epsilon_i, i \in N$. as long as the $\epsilon_i, i \in N$ are independent and $U_0$ is deterministic

The random consideration set (RCS) model of \cite{manzini2014stochastic} has, by definition, independent value gaps because in the RCS model consumers are assumed to have {\em independent attention} probabilities, where the attention probabilities correspond to our last choice probabilities.  In the RCS model, consumers first identify the set $C(S) = \{i \in S: B_i = 1\}$ and then select a product in $C(S)$ according to a fixed preference ordering $\succ$.  Thus, for any $S \subset N$ the expected revenue under the \cite{manzini2014stochastic} model is given by 
$$R_{\succ}(S) = \sum_{i \in S} r_i \omega_i \Pi_{j \in S: j \succ i}(1-\omega_j).$$

 The assortment optimization problem for the RCS model was first considered by \citet{gallego2017attention}, where the authors show that the revenue-ordered heuristic achieves at least $0.5\cR^*_{\succ}$. By Proposition~\ref{prop:pi} we have that the revenue-order heuristic achieves at least $0.5\bar{\cR}$ strengthening their result.

The right-hand side of $\bar{\cR}$ in equation (\ref{eq:rcs}) corresponds to the \cite{manzini2014stochastic} model evaluated at $S = N$,  when the preference-order is high-to-low: $1 \succ 2 \succ \cdots \succ n$. So $\bar{\cR}  = \cR^*_{HL} = R_{HL}(N)$, where the subscript $HL$ denotes that preference for higher revenue products. In fact, for any $S \subset N$, $R_{HL}(S)$ denotes the maximal expected revenues for assortment $S$ among the class of DCMs with independent $B_i, i \in N$.

At the other extreme, consider the RCS model with preference ordering $n \succ n-1 \succ \cdots \succ 1$. Then $R_{LH}(S) =  \sum_{i \in S} r_i \omega_i \Pi_{j \in S: j > i}(1-\omega_j)$ is the expected revenue under assortment $S$ for the RCS model where the preference now is for lower revenue products. Our proof of Proposition~\ref{prop:pi} implies that 
$R_{LH}(S) \geq 0.5 R_{HL}(S)$ for all $S \subset N$, and in particular $R_{LH}(N) \geq 0.5R_{HL}(N)$, a fact that is non-trivial to establish without Proposition~\ref{prop:pi}, as can be seen in \cite{LU2024107070}.





\subsection{The MNL Model.}\label{sec:MNL}

We remark that for the MNL the value gaps $U_i - U_0, i \in N$ are dependent, through $U_0$, so we cannot appeal to the basic prophet inequality for independent $B_i, i \in N$. Since 
$\cM_v(0,[i-1])- \cM_v(0,[i]) =  \cM_v(0, [i-1])\cdot \cM_v(i,[i])$
for the MNL, by Theorem~\ref{thm:fcr}
\begin{equation}
\label{eq:rmnl}
\bar{\cR}_v   =  \sum_{i \in N}r_i\cM_v(0, [i-1])\cdot\cM_v(i,[i]).
\end{equation}

\begin{theorem}\label{thm:mnl}
The prophet inequality holds for the MNL model. 
Moreover, the bound is tight. 
\end{theorem}

To see how a firm may segment consumers under the MNL model, consider the case of $n = 2$ and notice that under all preference orderings, except $2 \succ 0 \succ 1$, it is optimal to offer assortment $\{1\}$. Thus, if a firm can identify consumers with preference ordering $2 \succ 0 \succ 1$, then the firm can offer them assortment $\{2\}$ and offer assortment $\{1\}$ to everyone else. In this way the p-TAOP firm can earn as much as the clairvoyant firm. 

We remark that the tight bound mentioned in Theorem~\ref{thm:mnl} requires very different attraction values. In what follows we consider the case where the $v_i = 1, i \in N$. This situation arises, approximately, when the underlying Gumbel random variables have a very large variance. Indeed $v_i = \exp(u_i/\beta)$ where $u_i$ is the utility of product $i$ and $\beta$ is the scale parameter of the Gumbel distribution. The variance of the $\epsilon_i$  is proportional to $\beta^2$, so as $\beta \rightarrow \infty$,  $v_i \rightarrow 1$ for all $i \in N$. The following theorem provides a sharper result that suggests that for high-variance models the benefits of personalization under the MNL are even smaller. 
\begin{theorem}
\label{thm:bigbeta}
$$\cR^*_e \leq \bar{\cR}  \leq 1.5\cR^*_e.$$
\end{theorem}

\subsection{The Generalized Attraction Model}\label{sec:GAM}

The generalized attraction model (GAM) has $2n$ parameters; for each $i \in N$ there is a positive attraction value $v_i > 0$ and a shadow attraction value $w_i \in [0, v_i]$. The choice probabilities are given by 
$$\cP(i,S) = \frac{v_i}{1 + \sum_{i \in S}v_i + \sum_{i \in N\setminus S}w_i}, i \in S,$$
with $\cP(0,S) = 1 - \sum_{i \in S}\cP(i,S)$. The reader can verify that if $w_i = 0, i \in N$ then the GAM reduces to the MNL, and if $w_i = v_i, i \in N$ then the GAM reduces to the independent demand model. 

Basic algebra reveals that 
$$\cP(0,[i-1]) - \cP(0,[i]) = \frac{v_i\cP(0,[i]) + w_i(1- \cP(0,[i]))}{1 + \sum_{j < i}v_j + \sum_{j \geq i}w_j},$$
so by Thoerem~\ref{thm:fcr} we obtain

$$\bar{\cR} = \sum_{i \in N}r_i \frac{v_i\cP(0,[i]) + w_i(1- \cP(0,[i]))}{1 + \sum_{j < i}v_j + \sum_{j \geq  i}w_j}.$$

\begin{theorem}
\label{thm:GAM}
The prophet inequality holds for the GAM model. 
Moreover, the bound is tight. 
\end{theorem}

 
\subsection{The Latent Class MNL.} \label{sec:LCMNL}

The reader may wonder whether a prophet inequality will extend to personalized assortments when each type is governed by an MNL.  The answer is yes! If the firm can observe the type of each arriving consumer and offer them a personalized assortment the p-TAOP firm earns expected revenue  
$$\cR^p = \sum_{j \in M} \theta_j \cR^*_j,$$
where $\cR^*_j$ is the optimal expected revenue for type $j$ consumers. On the other hand, the clairvoyant firm earns 
$\bar{\cR} := \sum_{j \in M} \theta_j \bar{\cR}_{j}$, where  $\bar{\cR}_{j} := \sum_{i =1}^n r_i \cM_{j}(0,[i-1]) \cM_{j}(i, [i])$ is the expected profit of the clairvoyant firm for type $j$ consumers. 


\begin{corollary}\label{prop:personalized_by_segment}If each market segment is governed by an MNL model,
$$\cR^p \leq \bar{\cR}  \leq 2 \cR^p.$$
\end{corollary}
Corollary~\ref{prop:personalized_by_segment} asserts that under a LC-MNL model, a clairvoyant firm cannot make more than two times the optimal expected revenue of a p-TAOP firm that can customize an assortment for each type.

This, however, does not tell us how much more a clairvoyant firm can make relative to a firm that does not personalize assortments. In Theorem~\ref{thm:ubn} we show that it is possible to construct instances where the sequence of inequalities  
$\cR^p \leq \bar{\cR} \leq n \cR^o \leq n \cR^*$
are all arbitrarily tight. The instances constructed are over preference orderings, but since the class of DCMs governed by preference orderings are equivalent to the class of RUMs, and all RUMS can be approximated by LC-MNL models the bounds apply also to the LC-MNL model.

For non-personalized assortments the firm attempts to maximize $R(S) := \sum_{j \in M} \theta_j R_{j}(S)$ over $S \subset N$. This problem is well known to be NP-hard, see \cite{rusmevichientong2014assortment}.  We can write the last-choice probabilities for the LC-MNL as:
 $\omega_i = \cP(i,\{i\}) = \sum_{j \in M} \theta_j \cM_{j}(i,\{i\}) = \sum_{j \in M} \theta_j \omega_{ij}, i \in N$,
 where $\omega_{ij} = v_{ij}/(1+ v_{ij})$. We can then define the vector $\omega$ with components $\omega_i, i \in N$, and for this vector we compute $\cR^*_{\omega}$ and $S^*_{\omega}$. Condition (\ref{eq:condphi}) for $\phi$ is equivalent to
 \begin{equation}
 \label{eq:condlc}
\sum_{j \in M}\theta_j (v_{ij} - \phi \omega_i) \cM_{j}(0,S^*_{\phi \omega}) \geq 0~~~\forall i \in S^*_{ \phi \omega}.
\end{equation}
and can be checked efficiently as it merely requires computing $S^*_{ \phi \omega}$ and $O(n)$ operations to verify whether or not  (\ref{eq:condlc}) holds.


Numerical results in \S\ref{sec:exp} indicate that condition (\ref{eq:condlc}) holds  for $\phi = 1$ in virtually all of our experiments, so even though the clairvoyant firm can make up to $n$ times more than the TAOP firm, it never happened in the random instances generated in our studies.

\subsection{The Markov Chain Model}\label{sec:Markov}

Under the Markov Chain discrete choice model \citep{blanchet2016markov} an arriving consumer draws his or her first choice product and then evolve according to the transition matrix $\rho$, until absorption in $S_+$ where $S$ is the assortment offered by the firm.

From the proof of Theorem \ref{thm:ubn}, we know that there are instances of the MC model where $\bar{\cR}/\cR^*$, and $\cR^p/\cR^*$ are arbitrarily close to $n$. Personalized assortments, in this example, required knowing the consumer type, which was characterized by the highest index product preferred to the outside alternative. Given the specific form of the instance, this was tantamount to knowing the preference ordering of each arriving consumer. 

At the other extreme, we consider the case where the firm can only elicit the first-choice product for each arriving consumer. This amounts to finding out which is the first product traversed in the random walk procedure of the Markov chain choice model. Personalization by the first choice can be an effective mechanism for some choice models. The reader may wonder whether knowing the first choice may also be effective for the MC model. Surprisingly, we find that personalization based on first choice does not help at all under the MC model!  

\begin{proposition}
\label{prop:dcm}
    Knowing the first-choice of arriving consumers is not beneficial for the MC model. More precisely, under the Markov chain choice model, if $S^*$ is an optimal assortment for customers whose first choice is $i \in N$, then $S^*$ is also an optimal assortment for customers whose first choice is any other $j \in N$.
\end{proposition}

One may wonder whether personalization based on a deeper knowledge of the consumers' preference may yield higher benefits. We have seen that, if the firm has depth one knowldge (first choice) then personalization can make only $1$ times $\cR^*$, so there is no gain. On the other hand, if the firm has knowledge of depth $n$ (the entire preference ordering) then it can make up to $n$ times $\cR^*$. These observations begs the question of whether knowledge of depth $k$ under the MC choice model can lead to earning up to $k\cR^*$. We leave this as an open question for further research.

\subsection{The Nested Logit Model}\label{sec:NL}

In the nested logit (NL) model, the set $N$ is partitioned into a collection of mutually exclusive and collectively exhaustive subsets.  More precisely, $N = \cup_{k \in K}N_k$,  $N_k \neq \emptyset, k \in K$, and $N_k \cap N_l = \emptyset$ for all $k \neq l$. Each nest $k \in K$ has a dissimilarity parameter $\gamma_k \geq 0$. Moreover, associated with each alternative $i \in N_+$ there is a positive attraction value $v_i > 0$.

Let $S_k \subset N_k$ be the set of products offered in nest $k \in K$. Then $S = \cup_{k \in K}S_k$ is the offered assortment. Conversely, for any assortment $S \subset N$, the subset offered to nest $k$ is given by $S_k = S \cap N_k, k \in K$. Given an assortment $S$, the probability that nest $k$ is selected is given by 
$$Q(k,S) := \frac{V_k(S_k)}{v_0 + \sum_{l \in K}V_l(S_l)}~~~~k \in K,$$
where fore for any $S \subset N$,  $V_k(S) := \left(\sum_{i \in S \cap N_k}v_i\right)^{\gamma_k}$, 
and therefore $V_k(S) = V_k(S_k)$ on account of $S_k = S \cap N_k$. 

The choice probabilities under the NL model are given by 
$$\cP(i,S) := Q(k,S) \frac{v_i}{\sum_{j \in S_k}v_j}~~~\forall~~~i \in S_k, k \in K,$$
so conditioned on nest $k$ being selected, the probabilities are apportioned in proportion to their attraction values,  so once a nest is selected a purchase must be made within the nest. Clearly,
$\cP(0,S) = Q(0,S) := \frac{v_0}{v_0 + \sum_{l \in K}V_l(S_l)}$.

If nest $k$ is selected then it must be that $S_k \neq \emptyset$. The expected revenue from nest $k$, conditioned on nest $k$ being selected, is given by
$R_k(S) :=  \sum_{i \in S \cap N_k} r_i\frac{v_i}{\sum_{j \in S \cap N_k}v_j}$
so $R_k(S) = R_k(S_k)$. 
We extend the definition of $R_k$ by setting $R_k(S) = 0$ whenever $S_k = \emptyset$.  
 We can now write the expected revenue for assortment $S \subset N$ as $R(S) := \sum_{k \in K}R_k(S_k)Q(k,S)$. Simple algebraic manipulations lead to the following expression:
\bean
R(S) & =  &  \sum_{k \in K}[R_k(S_k) - R(S)]\frac{Q(k,S)}{Q(0,S)} =   \sum_{k \in K}[R_k(S_k) - R(S)]\frac{V_k(S_k)}{v_0}.
\eean

Clearly $\tau = R(S)$ is a root of  the equation $\tau = \sum_{k \in K}[R_k(S_k) - \tau]\frac{V_k(S_k)}{v_0}$, and this root is unique because the left hand side is strictly increasing in $\tau$ while the right hand side is decreasing in $\tau$.

The formula above suggests an  idea to find an assortment $T$ with higher expected revenue when $L(S) := \{k \in K: [R_k(S) - R(S)]V_k(S_k) < 0\} \neq \emptyset$. The following results are refinements of earlier results in \cite{davis2014assortment}. 

 \begin{lemma} 
 \label{lem:nlm}
    If $L(S)$ is non-empty, construct $T$ by setting $T_k = S_k$ for all $k \notin L(S)$, and $T_k = \emptyset$ for all $k \in L(S)$. Then $R(T) > R(S)$. Moreover, if $L(S) = \emptyset$ then $R(S)$ is the unique root of the equation $\tau = \sum_{k \in K}[R_k(S_k) - \tau]_+V_k(S_k)/v_0$.  
 \end{lemma}

\begin{theorem}
\label{thm:nlm}
    Suppose that $S^* \in \arg \max_{S \subset N}R(S)$ and $\cR^* = R(S^*)$. Then for every $k \in K$, either $R_k(S^*_k) \geq \cR^*$ or $S^*_k = \emptyset$. 
    Moreover, $\cR^* = R(S^*)$ is the unique root of the equation 
    $\tau = \sum_{k \in K}[R_k(S^*_k) - \tau]_+ \frac{V_k(S^*_k)}{v_0}$.
\end{theorem}

Our results so far make no assumptions about the dissimilarity parameters. The special case where $\gamma_k \leq 1,  k \in K$ results in the standard NL model which is known to be a RUM. Relaxing this constrained results in a non-standard NL model. Allowing for $\gamma_k > 1$ can result in a better fit to data, as explained in \cite{davis2014assortment} and references therein. However, R1 may fail when $\gamma_k >1$ for one or more nests. Consequently, $\bar{\cR}$ as computed by formula (\ref{eq:fcr}) may fail to be an upper bound on $\cR^*$. This happens because of the positive synergy among the products in the same nest, when $\gamma_k > 1$, may make it sub-optimal to offer naked products. Although the general NL model may fail to be regular, it is still {\em cross-nest regular} in the sense that adding product to a nest reduces the probability of selecting products in other nests. 
 
The fact that R1 may fail to hold suggests that a clairvoyant firm needs to be more sophisticated and determine the best assortment to offer rather than the best individual product to offer. Thus, a full clairvoyant firm must be able to correctly guess the highest revenue product that an arriving consumer is willing to purchase, {\em and} the assortment that needs to be offered to elicit such sales. This is in contrast to the case where Assumption R holds where we know that it is enough to offer a product by itself. The problem with such a clairvoyant firm is that  even if two consumers are willing to buy product $i$, they may require different assortments to trigger the purchase and this makes the formulation of the expected revenue of the fully clairvoyant firm intractable.

To avoid these complications we will consider a nest-clairvoyant firm that can tell for any assortment $S$ which nests the consumer would be willing to buy from when {\em only} products in that nest are offered. More precisely, the nest-clairvoyant firm can tell if nest $k$ would be selected under assortment $S_k = S \cap N_k$ for all $k \in K$. The expected revenue from offering this nest is $R(S_k) = Q(k,S_k)R_k(S_k)$ for all $k \in K$. We can think of $R(S_k)$ as the expected value of the random variable $X_k(S_k) = R_k(S_k)B_k(S_k)$ where $B_k(S_k)$ is a Bernoulli random variable that takes value one with probability $Q(k,S_k)$ and value $0$ otherwise. 

Similarly, we define $B_k(S)$ as a Bernoulli random variable that takes value one with probability $Q(k,S) \leq Q(k,S_k)$, and value zero otherwise. We assume that R1 holds at the nest level, meaning that if $B_k(S) = 1$ for any $S$ such that $S_k = S \cap N_k$, then $B_k(S_k) = 1$. In other words, if nest $k$ is selected under the assortment $S$, then nest $k$ is also selected under the restricted assortment $S_k = S\cap N_k$.  We next argue that under this mild assumption the nest-clairvoyant firm can make at least as much revenue as the TAOP firm. To see this, suppose that the TAOP firm offers assortment $S$ and the consumer selects nest $l \in K$. Then nest $l$ will also be selected under assortment $S_l$ and the nest-clairvoyant firm can offer assortment $S_l$ and earn what the TAOP firm earns. However, the nest-clairvoyant also knows if there is a different nest, say $k$, with $R_k(S_k) > R_l(S_l)$ and $B_k(S_k) = 1$. Since the nest-clairvoyant firm can select the best nest to offer it can earn at least as much as the TAOP firm and often strictly more. We have shown that under our mild assumption that R1 holds at the nest level  
$\bar{\cR}(S) := E[\max_{k \in K}X_k(S_k)] \geq R(S)~~~\forall ~~S\subset N$. Optimizing over $S$ on both sides we obtain
$$\bar{\cR}  := \max_{S \subset N}E[\max_{k \in K}X_k(S_k)] \geq \max_{S \subset N}R(S) = \cR^*.$$

Here $\bar{\cR}$ denotes the maximum expected revenue that can be earned by the nest-clairvoyant firm, while $\cR^*$ is the expected revenue of the TAOP firm. We are now ready to show that a prophet inequality holds relative to the nest-clairvoyant firm.

\begin{theorem} For the NL model the nest-clairvoyant firm can earn at most $2\cR^*$, so $\bar{\cR} \leq 2 \cR^*$.
\label{thm:nlsc}
\end{theorem}

Remark 1: The nested-clairvoyant firm can do better than any effort to personalize assortments that does not have more information than the nested-clairvoyant firm. For example, the firm may have information about how consumers may be biased in selecting the nests with the aggregate demand over all consumers being the NL model. The firm can then personalized assortments based on these biases. However, the nest-clairvoyant firm will always manage to make as much money because it can determine which nests the consumer is willing to buy and then concentrate on the best nest. Theorem~\ref{thm:nlsc} therefore tells us that any efforts to personalize with less knowledge will result in at most $\bar{\cR}$ and therefore at most $2\cR^*$. 

Remark 2: Theorem~\ref{thm:nlsc} suggests that we can upper bound the expected benefit from personalization via the ratio of $\bar{\cR}/\cR^*$. However, $\cR^*$ is NP-hard to compute, and most likely so is $\bar{\cR}$. Therefore, Theorem~\ref{thm:nlsc}, is currently of a theoretical nature, telling us that personalization satisfies the prophet inequality relative to the nest-clairvoyant firm even if there is no efficient algorithm to compute the ratio.

Remark 3: The reader may wonder if the prophet inequality for the nest-clairvoyant firm would hold if we allow a no-purchase alternative at the nest level. This would require setting  
$V_k(S) := (v_{0k} + \sum_{j \in S_k}v_j)^{\gamma_k}$, and $R_k(S) := \sum_{i \in S_k} r_i\frac{v_i}{v_{0k} + \sum_{j \in S_k}v_j}$. The main difficulty is that the formula for $\cP(0,S)$ becomes more complicated given that we must now account for the possibility that the no-purchase alternative is selected at the nest stage. Nevertheless, it is plausible that the ideas presented here may be extended to incorporate this more nuanced model. We leave it to interested readers to pursue this avenue.

\section{Experiments}
\label{sec:exp}
We next present the results of computational experiments under the LC-MNL model to study the relative performance of traditional assortment optimization, personalized assortment optimization and the clairvoyant firm. We present these results as percentages of the revenue obtained for a TAOP-firm that employs the revenue-ordered assortment heuristic.

\begin{figure}[ht]
\centering
\includegraphics[scale=0.52]{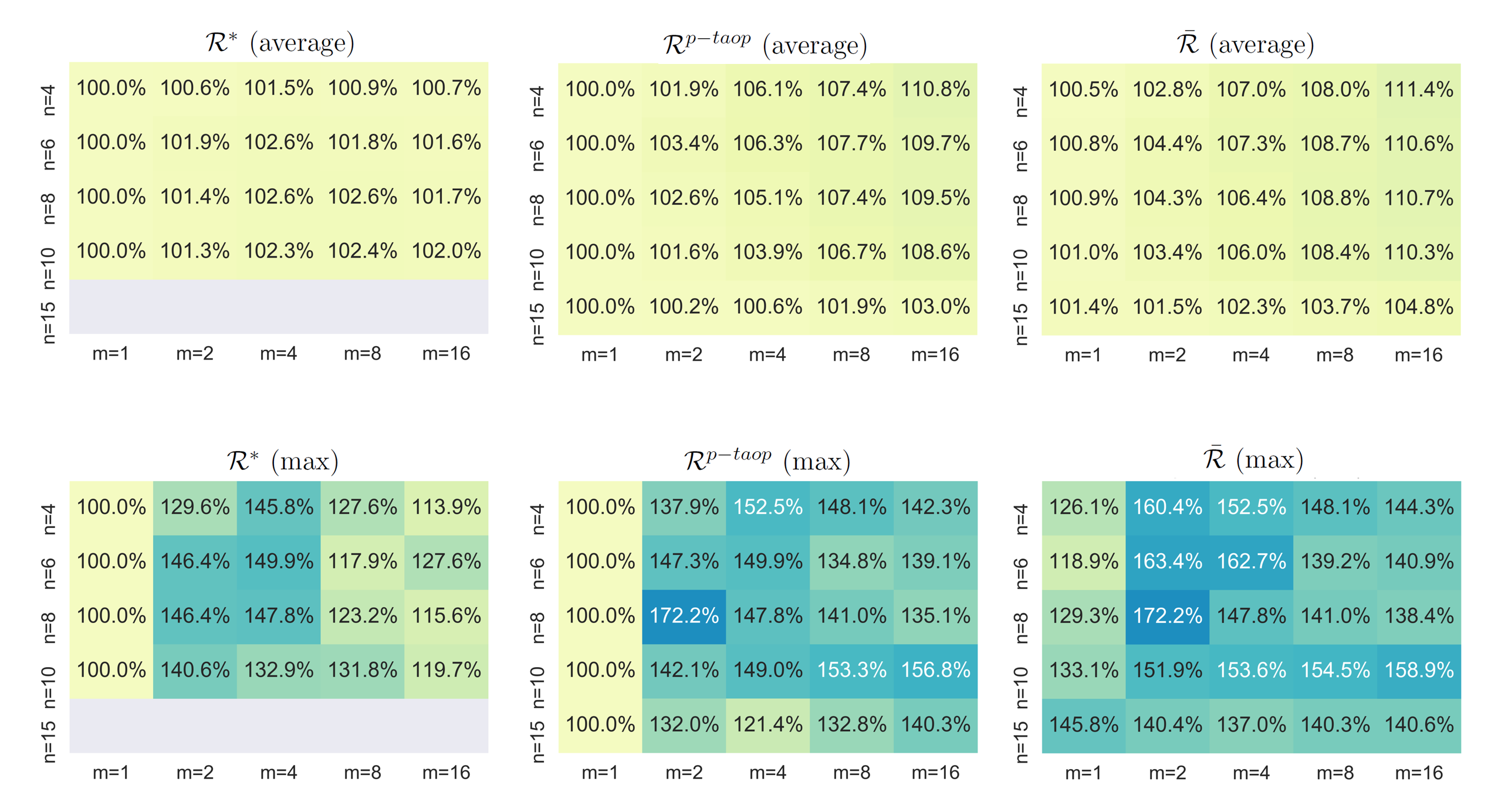}
\caption{Scenario 1: $\beta=0.02$. Performance of TAOP, personalized TAOP and a clairvoyant as a percentage of revenue-ordered profits under the LC-MNL model. For each value of $n$ and $m$, we performed 300 experiments.}
\label{fig_lc-mnl-theta_0.02}
\end{figure}

\begin{figure}[ht]
\centering
\includegraphics[scale=0.52]{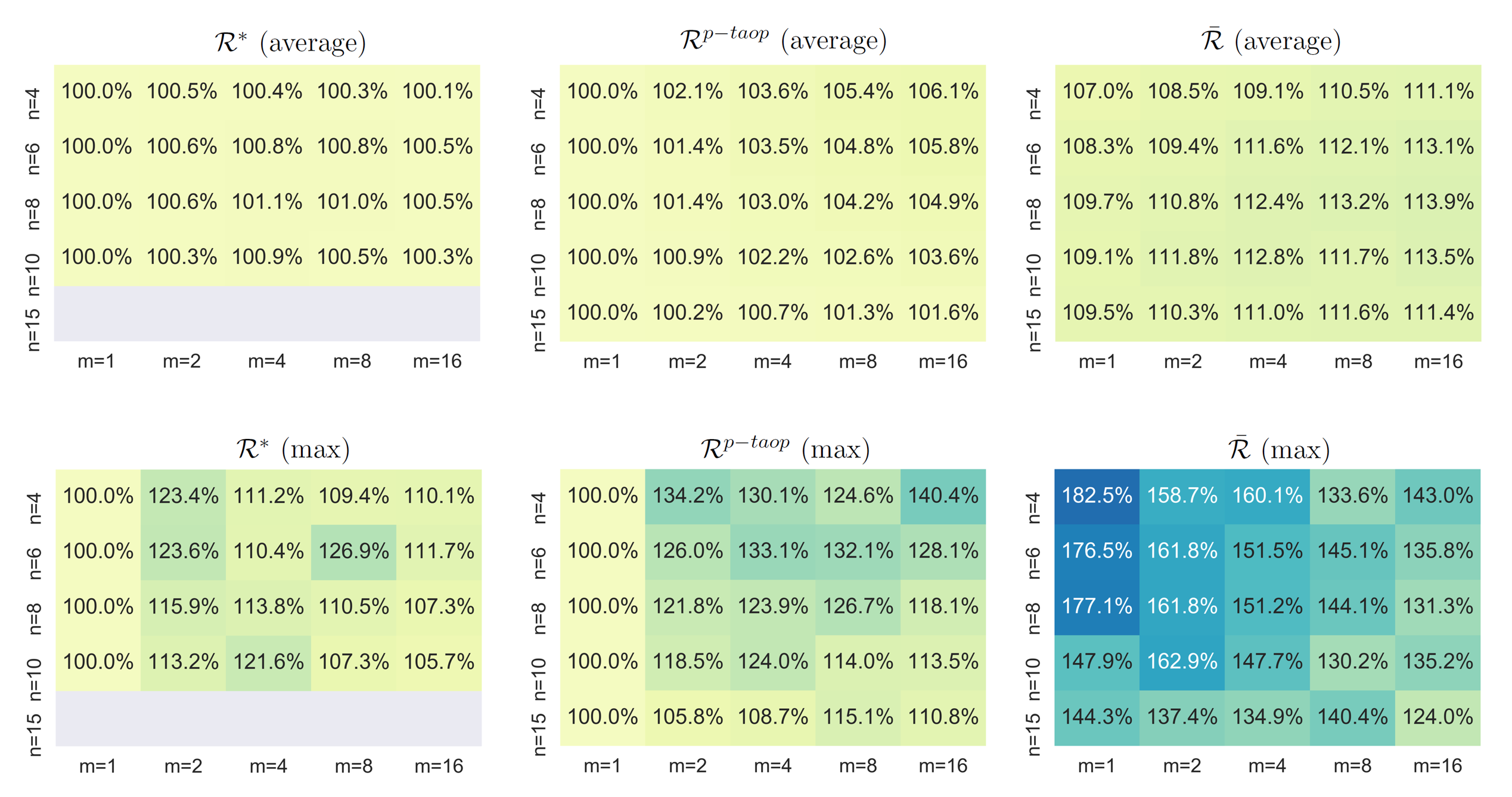}
\caption{Scenario 2: $\beta=0.2.$ Performance of TAOP, personalized TAOP and a clairvoyant as a percentage of revenue-ordered profits under the LC-MNL model. For each value of $n$ and $m$, we performed 300 experiments.}
\label{fig_lc-mnl-theta_0.2}
\end{figure}

\begin{figure}[ht]
\centering
\includegraphics[scale=0.52]{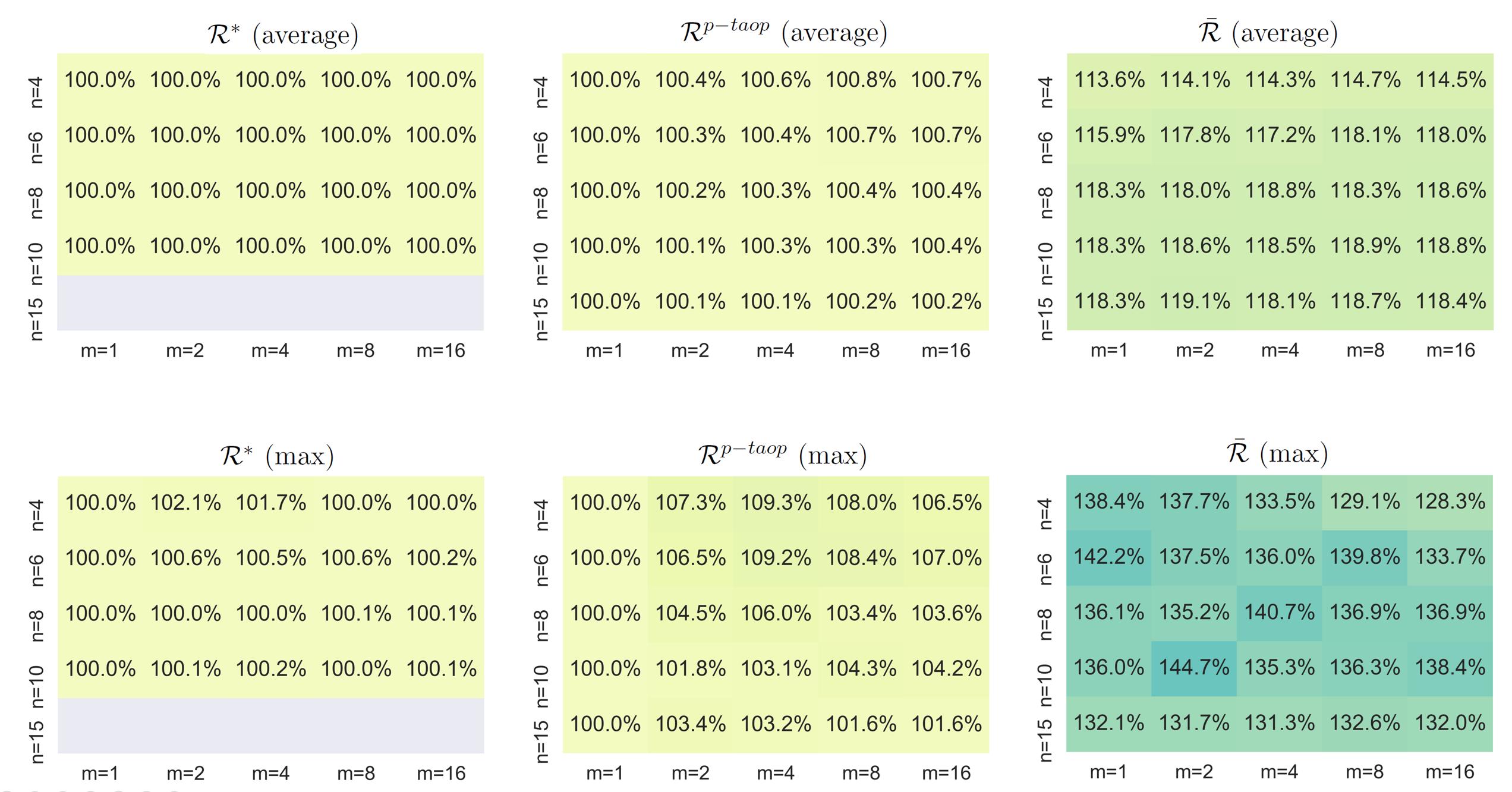}
\caption{Scenario 3: $\beta=2$. Performance of TAOP, personalized TAOP and a clairvoyant as a percentage of revenue-ordered profits under the LC-MNL model. For each value of $n$ and $m$, we performed 300 experiments.}
\label{fig_lc-mnl-theta_1}
\end{figure}

\begin{figure}[ht]
\centering
\includegraphics[scale=0.52]{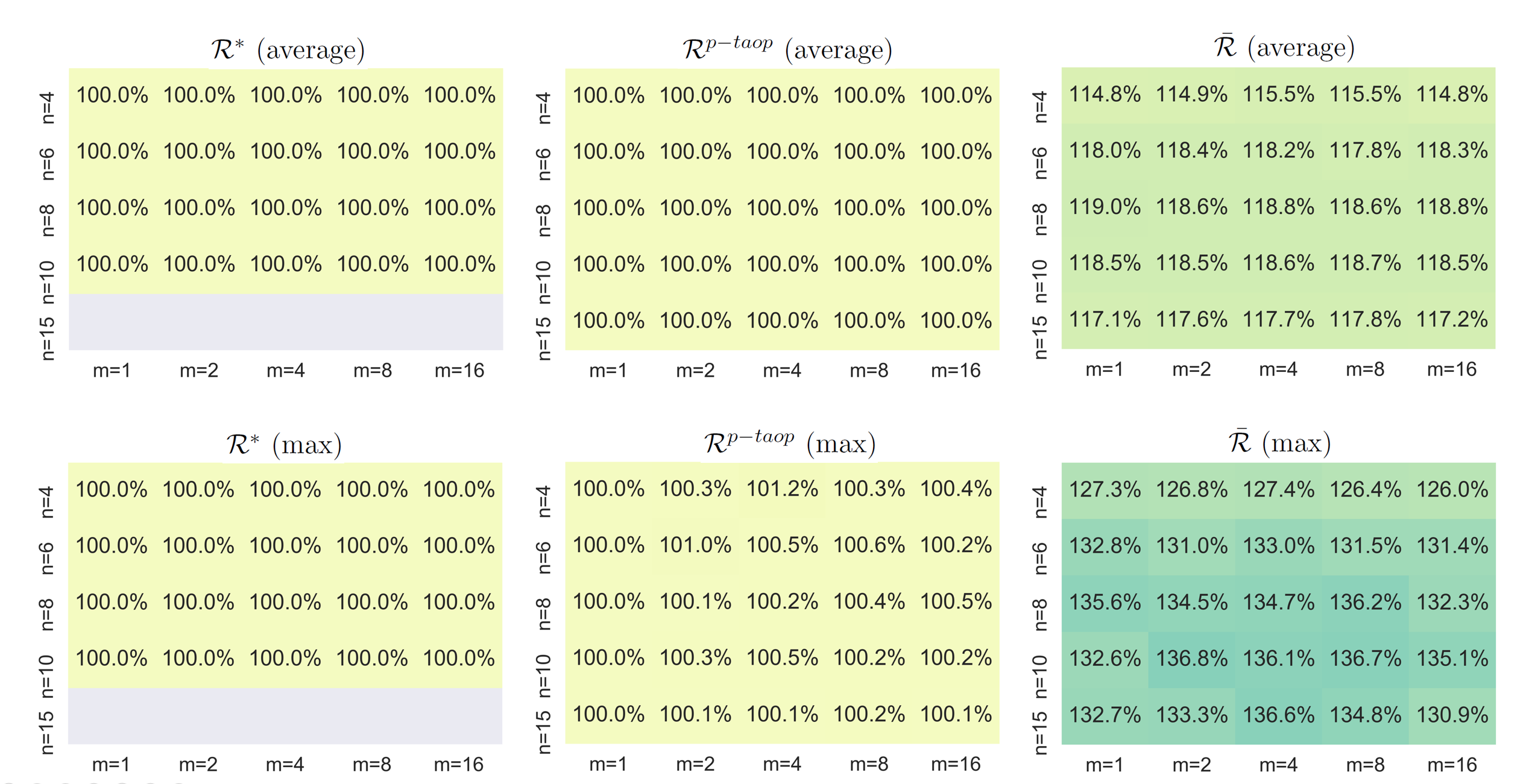}
\caption{Scenario 4: $\beta=20$. Performance of TAOP, personalized TAOP and a clairvoyant as a percentage of revenue-ordered profits under the LC-MNL model. For each value of $n$ and $m$, we performed 300 experiments.}
\label{fig_lc-mnl-theta_20}
\end{figure}

The instances for the numerical study follow a well accepted procedure to generate random instances first proposed by \cite{rusmevichientong2014assortment}. The utility of $i \in N$ to type $j \in M$ is modeled as $U_{ij}= u_{ij} + \epsilon_{ij}$ where  $u_{ij}$ is the deterministic part of the utility and $\epsilon_{ij}, i \in N_+$ are standard Gumbel random variables with mean zero and variance $\pi^2/6$, corresponding to scale parameter 1 and location parameter $-\gamma$ where $\gamma$ is the Euler's constant. By setting $u_{ij} = a_{ij}/\beta$ for a parameter $\beta > 0$ we can control the variance of the utilities. Indeed, by multiplying all the utilities by $\beta$, the variance of each Gumbel changes to $\pi^2\beta^2/6$. Notice that the exponentiated utilities are given by $v_i = \exp(a_{ij}/\beta)$. In our experiments, we fixed the lowest revenue to $1$ and the highest revenue to $10$, and selected the revenues of the rest of the products uniformly between $1$ and $10$.  For each combination of $(m,n)$ we generated 300 random instances. For each instance, the $a_{ij}, i \in N_+$ are chosen randomly following \cite{rusmevichientong2014assortment} \footnote{Specifically, $a_{ij}$ (which represents the nominal utility of product $i$ in segment $j$ in their paper), is defined as zero in case $i=0$, otherwise $a_{ij}:=\ln((1-\sigma_i)\ell_{ij}/n)$ with probability $p=0.5$ and $a_{ij}:= \ln((1+\sigma_i)\ell_{ij}/n)$ in the other case. The values $\ell_{ij}$ and $\sigma_i$ are realizations from a uniform distribution $(0,10]$ and $(0,1]$ respectively.} and  present results for four different values of $\beta$: 0.02 (fig. \ref{fig_lc-mnl-theta_0.02}); 0.2 (fig. \ref{fig_lc-mnl-theta_0.2}); 2 (fig. \ref{fig_lc-mnl-theta_1}), and 20 (fig. \ref{fig_lc-mnl-theta_20}). For each of those four scenarios, we calculate the optimal revenue obtained under TAOP; personalized TAOP (p-TAOP)\footnote{This is the optimal revenue obtained when the firm can offer a personalized assortment to each consumer type. Namely, $R^{p-TAOP}:= \sum_{j=1}^m w_jR_j(S^*_j)$ where $S^*_j$ is an optimal assortment to segment $j$, $w_j$ is the segment $j$ weight and $R_j(S)$ is the revenue obtained from segment $j$ when offered assortment $S$. See Section \ref{sec:p-taop}}; and the clairvoyant firm ($\bar{\cR}$) as a percentage of the revenue obtained using revenue-ordered assortments heuristic under traditional assortment optimization. Each figure reports the average and maximum percentage across the 300 instances.


Our computational experiments suggest that the revenue-ordered heuristic performs relatively well even against a clairvoyant firm. Indeed, from our experiments we see that the expected revenues of the clairvoyant firm are, on average, between 0.5\% and 19\% higher than the expected revenues under the revenue-ordered heuristic.  On the other hand, the personalized heuristic is on average 0\% to 10.8\% higher than the expected revenue of the revenue-ordered heuristic. 


A much clearer picture emerges when we study the numerical results for different values of $\beta$ which is a proxy for the variance of the utilities $U_{ij}, i \in N, j \in M$. For $\beta = 0.02$ (fig \ref{fig_lc-mnl-theta_0.02}) there can be a significant gap between $\cR^p$ and $\cR^*$,  but a relatively small gap between the p-TAOP and the clairvoyant firm. Exploring the instances where the p-TAOP firm makes significant gains over the TAOP firm, we discover that these instances have relatively high heterogeneity in the mean utilities as well as low variance. This makes sense because with heterogeneity and low variance a p-TAOP firm can guess with high probability the product(s) the consumer is likely to buy. This also suggests that in this instances the gap between the p-TAOP and the clairvoyant firm should be small! This is indeed confirmed by instance $(m, n) = (16, 4)$ where on average the p-TAOP makes 10.8\% more than the revenue-ordered heuristic while the clairvoyant firm makes  11.4\%  more than the revenue-ordered heuristic. 

We can make these insights more formal by looking at what happens as $\beta \downarrow 0$.  Suppose that the random utilities $U_{ij} = u_{ij} + \epsilon_{ij}, i \in N_+, j \in M$ are such that for every $j$ there the $u_{ij}$s are well spread, and the variance of the $\epsilon_{ij}, i \in N$ are small for every $j \in M$. Then, for each customer segment the model converges to a maximum utility model. Consequently, the p-TAOP firm can personalize assortments based on the purchase probabilities $\cP_j(i, \{i\}) = \omega_{ij}$ which in the limit are either $0$ or $1$. Thus, as $\beta \downarrow 0$, the ability of the p-TAOP firm approaches the ability of the clairvoyant firm. This justifies why for small $\beta$ we see a small gap between the expected revenue of a p-TAOP firm and the clairvoyant firm. In this case, the p-TAOP and the clairvoyant firm have a significant advantage relative to the TAOP firm as long as there is heterogeneity among the different consumer types, and also heterogeneity in the $r_i, i \in N$.  

On the other hand, when $\beta  = 20$ (Fig \ref{fig_lc-mnl-theta_20}) the optimal revenue under the TAOP and the p-TAOP are almost indistinguishable from each other and from the expected revenue of the revenue-ordered heuristic.  Indeed, the p-TAOP makes at most 0.8\% on average more than the revenue-ordered heuristic over all the $(m,n)$ combinations. Notice also that there is a significant gap between the p-TAOP and the clairvoyant firm, but the ratio is never larger than 1.5.  

We can also make these insights more formal. First, if the variances are large then the exponetiated utilities $v_{ij} = \exp(u_{ij}/\beta) \approx 1$. So the DCM for each $j \in M$ becomes close to an MNL with attraction vector $v = e$. In this case, the LC-MNL itself converges to an MNL with $v = e$. It is clear that in the limit there is no advantage of personalization and therefore $\cR^{p} = \cR^* = \cR^0$. On the other hand, we see numerically that the clairvoyant firm has a decisive advantage with earnings up to 1.447 times that of $\cR^o$.  In Theorem~\ref{thm:bigbeta} we show that the clairvoyant firm can earn only up to 1.5 more than a TAOP firm, lending theoretical support to our numerical findings.
In the experiments described above, the variance in the product's utility which is controlled by the parameter $\beta$ is the same for all consumer segments. We ran similar experiments for instances where the value of $beta$ is different for the segments and the results are quite similar to the ones presented.

\section{Examples of Personalized Assortments}\label{sec:examples}

In this section we provide two examples of personalized assortments. The first example explores the possibility of personalizing based on eliciting first choice information from arriving consumers. We provide an example where personalization based on first choice results in a significant increase in revenues. On the negative side we have shown, see Proposition~\ref{prop:dcm}, that personalization based on first choice does not work for the class of Markov Chain (MC) DCMs. This implies that personalization based on first-choice will not bring any extra profits for the firm if the underlying DCM is an MNL, a GAM, or a RCS model as all of these are special instances of the MCM. 

We also explore examples where it may be possible to personalize assortments based on consumers' covariates. In this case, the firm observes some characteristics of the consumer and personalize based on them, perhaps after some clustering.

\subsection{Personalizing Assortments Based on First Choice}

An intriguing possibility is to elicit from each arriving consumer their preferred, or \textit{first choice} (FC), product in $N$. Given that $i \in N$ is the first choice, the firm offers an assortment, say $S^*(i)$, that optimizes the firm's expected revenue for consumers whose first choice is product $i$.  This requires the decision maker to know the conditional probabilities $\cP(j,S|FC = i)$ for all $j \in S_+, S \subset N$, for all $i \in N$.

As an example, suppose that $N = \{1,2,3\}$, and that an arriving consumer is equally likely to be of type $j \in  M = \{1,2,3\}$. Suppose  and that preference ordering of the three types are respectfully $1 \succ 2 \succ 0$, $2 \succ 0 \succ 1$, and $3 \succ 2 \succ 1 \succ 0$. It is easy to see that if $r_2(\lambda_2 + \lambda_3) \leq r_1\lambda_3$, then $\cR^* = R(\{1\}) = r_1(\lambda_1 + \lambda_3)$. A firm that can elicit the first-choice from consumers can personalize assortments and offer assortment $\{2\}$ for consumers of type 2, and offer assortment $\{1\}$ otherwise. This results in $\cR^p = \bar{\cR} = \cR^* + r_2 \lambda_2$. In this example we can make $\bar{\cR}/\cR^*$ arbitrarily close to 2 by setting $r_2 = r_1 \lambda_3/(\lambda_2 + \lambda_3)$, $\lambda_1 = \epsilon^2$, $\lambda_3 = \epsilon$, and letting $\epsilon$ go to zero.

\subsection{Personalizing by consumer covariates}

Suppose that each arriving consumer has a vector $x \in \Gamma$ of covariates known to the firm. Then the covariate space $\Gamma$ can be partitioned into disjoint clusters $\Gamma_j, j \in M$, whose choice model can be approximated by a DCM. One possibility is to model the DCM for each type as MNL, as we have done in the numerical examples in \S~5. The drawback is that an MNL may not be flexible enough to correctly approximate the decisions of consumers in cluster $j$. An alternative is to use something more general such as a MC model, but that would require estimating $O(n^2)$ parameters for each $j \in M$. An intermediate possibility is to estimate DCM-$j$ by a parsimonious MC that requires estimating $O(n)$ parameters for each $j \in M$. A good candidate for this is the GAM, see \citep{gallego2015general}.

We present an example here to illustrate.  Recall that a GAM is specified by two vectors: the vector of attractions $v$, and the vector of shadow attractions $w$. 

Suppose that all arriving consumers have the same attraction vector $v$ for the products but that consumers can be classified based on the willingness or unwillingness to explore the offerings of competing retailers. At one extreme we have busy consumers that make a selection without bothering to see what other retailers are offering. At the other extreme we have consumers with time on their hands to explore the offerings of other retailers. To distinguish between these two categories, we call the former executive consumers and the latter household consumers. 

A simple model would have a common vector of attraction $v$ for all consumers, with $w = 0$ for executive consumers, and $w = v$ for household consumers. The following example illustrates how  knowing the consumer's type can lead to significant gains from  personalized assortments. Assume there are ten products. The vectors $r$ and $v$ are given in Table~\ref{tab:gamper}. We assume that half of the consumers are executive and the other half are household consumers. It is optimal to offer $S = \{1,2\}$ to executive consumers and to offer $S = N$ to household consumers. An optimal, non-personalized assortment is a compromise that offers $S = \{1,\cdots,7\}$. The expected gains from personalizing assortments is 19.12\%. If the attraction values are changed to $v_i = 100$ for all $i \geq 3$, then it is optimal to offer the same assortments to executive and household consumers, but now it is optimal to offer $S = \{1,2\}$ as the common non-personalized assortment. In this case the gains personalization are 52.38\%. It can be shown that even in the case of $n = 2$ it is possible to have up to 100\% gains in personalization by setting $v_1 = \epsilon$, $\theta = \epsilon^2$, $v_2$ large, and $r_2 = \theta \omega_1$ where $\theta$ is the fraction of executive consumers.


\begin{table}[]
    \centering
    \begin{tabular}{|c|c|c|c|c|c|c|c|c|c|c|}
    \hline
    $r$ & \$84.00 & \$73.50 & \$50.00 & \$49.00 & \$44.80 & \$31.50 & \$28.00 & \$23.20 & \$19.75 & \$18.30 \\
    \hline
    $v$ & 1.19584 & 3.23724 & 1.11864 & 2.29598 & 1.21726 & 3.19560 & 0.86152 & 0.41530 & 1.74140 & 0.67811 \\
    \hline
    \end{tabular}
    \vspace{0.1in}
    \caption{Revenues and Attraction Values for Executive and Household Consumers}
    \label{tab:gamper}
\end{table}



\section{Discussion}
\label{sec:disc}
We studied the limits of assortment personalization by considering the extreme case in which a firm is clairvoyant. While the clairvoyant firm can make up to $n$ times more than a TAOP firm over the class of Markov Chain models, it cannot make more than twice as much for a variety of commonly used DCMs. These include the MNL, the GAM, the Random Consideration Set Model. For the Nested Logit Model we have a prophet inequality relative to the nest-clairvoyant firm. In addition, we have also provide sufficient conditions that can be used to test whether or not the prophet inequality holds. Sharper inequalities also emerge for the LC-MNL model when the coefficient of variation of all the products is large, explaining why the revenue-ordered assortment perform so well in practice in such situations. Our computational results for the LC-MNL model support our theoretical results and show that the best revenue-ordered assortments does remarkably well even against a clairvoyant firm. We now discuss a few extensions to our model.



\subsection{Clairvoyant pricing} \label{sec:pricing}
Consider now a clairvoyant firm that observes the gross utilities $U_i, i \in N_+$ of each incoming consumer. How should such a clairvoyant firm set prices to maximize expected profits? Assuming zero marginal costs, for $n = 1$ it is  optimal to set $p_1$ as the largest non-negative price such that $U_1 - p_1 \geq U_0$.  This leads to $p_1 = (U_1 - U_0)^+$. Consequently, for the case of a single product the firm earns $E[(U_1-U_0)^+]$ in expectation. For multiple products, let $U_N : = \max_{i \in N}U_i$. Then it is easy to see that the firm should set $p_i = (U_N - U_0)^+$ for all $i\in N$, so the clairvoyant firm earns
$\bar{R} = E[(U_N-U_0)^+]$. 
On the other hand, a non-clairvoyant firm that is not allowed to personalize prices will earn expected profit $$\cR^* = \max_p \sum_{i \in N} p_i \cP(U_i - p_i \geq U_j -p_,~ j \in N, j \neq i, ~~U_i - p_i \geq U_0).$$
Clearly $\cR^* \leq \bar{\cR}$. As usual we seek bounds for the ratio of $\bar{\cR}/\cR^*$.

\begin{proposition}\label{prop:pricing_unbounded}
The ratio $\frac{\bar{\cR}}{\cR^*}$ can be arbitrarily large.
\end{proposition}

The next result shows that things are significantly better for the firm under the MNL model. 

\begin{proposition}
\label{prop:MNLprice}
For the MNL model, if both the traditional firm and the clairvoyant firm are free to select prices then, the ratio $\frac{\bar{\cR}}{\cR^*}$ is at most $e=\exp(1)$, and the bound is tight.
\end{proposition}

The result for the MNL readily extends to the LC-MNL problem if personalized pricing is allowed, so if $\cR^p$ is the expected profit from personalized pricing, then $\bar{\cR} \leq \exp(1) \cR^p$. Furthermore,
 we can obtain a worst case bound for $\cR^*$ relative to $\bar{\cR}$ that is $\exp(1)$ times larger than the worst-case bounds in \citet{gallego2021bounds} for $\cR^p$ relative to $\cR^*$.


\subsection{A joint assortment and customization problem}\label{sec:joint_assortment}
Recently, \cite{el2021joint} considered a joint assortment and customization problem under the LC-MNL model. This problem, called the \emph{Customized Assortment Problem} (CAP), consists of two stages. In the first stage, the firm needs to select a subset $T$ of at most $k$ products. In the second stage, the firm observes the consumer type $j \in M$ and chooses a personalized subset $S_j \subseteq T$ of products to offer. Thus, the CAP is the following optimization problem:

$$\mathcal{R}^*_{cap} = \max_{T \subseteq N, |T|\leq k} \sum_{j \in M}\theta_j \max_{S \subseteq T} \mathcal{R}_j(S)$$
where $\mathcal{R}_j(S)= \sum_{i \in S}\mathcal{M}_j(i,S)r_i$ denotes the expected revenue for segment $j$ when we offer assortment $S$.

\cite{el2021joint} proved that CAP is NP-hard\footnote{Finding an optimal assortment $T$ is the hard problem since the second stage assortment $S$ is simply a revenue-ordered assortment subset from $T$ which can be quickly computed.} and proposed a polynomial-time algorithm called \emph{Augmented Greedy} that guarantees at least a $\Omega(1/(\ln(m))$-fraction of the optimal revenue. More recently, \citet{udwani2021submodular} improved the revenue guarantees by constructing a $(0.5-\epsilon)$-approximation algorithm for the same problem.

A natural way to extend the CAP is to let the firm be a clairvoyant at the second stage so that it can customize the assortment offered to the specific individual rather than to the consumer type. The \emph{clairvoyant}-CAP is defined as follows:


$$\mathcal{R}^*_{clairvoyant-cap} = \max_{T \subseteq N, |T|\leq k} \sum_{j \in M}\theta_j \bar{\cR}_j(T)$$
where $\bar{\cR}_j(T)$ denotes the expected revenue obtained by a clairvoyant firm with universe of products $T$ that is faced by segment $j$ consumers.

Clearly, $\mathcal{R}^*_{cap} \leq \mathcal{R}^*_{clairvoyant-cap}$. Combining some of our clairvoyant results with results from \cite{el2021joint} and \citet{udwani2021submodular} it is straightforward to show the following propositions.

\begin{proposition}\label{prop:factor_2:joint}
$\mathcal{R}^*_{clairvoyant-cap} \leq 2 \mathcal{R}^*_{cap}$
\end{proposition}

\begin{proposition}\label{clairvoyant_CAP_np-hard}
\emph{Clairvoyant}-CAP is NP-hard.
\end{proposition}

Given the NP-hardness result for $\emph{clairvoyant}$-CAP, we are interested in approximation algorithms. We will consider algorithms to approximate $\emph{clairvoyant}-CAP$ that observe the consumer type but not the Gumbel noises associated to each specific consumer. The following two propositions directly follow from Proposition \ref{prop:factor_2:joint} and the revenues guarantees obtained by \citet{el2021joint} and \citet{udwani2021submodular}.

\begin{proposition}
The Augmented-Greedy algorithm \citep{el2021joint} provides an $\Omega(1/\ln(m))$-approximation to \emph{clairvoyant}-CAP.
\end{proposition}

\begin{proposition}\label{prop_0.5}
Algorithm 1 from \citet{udwani2021submodular} provides a $(0.25-\epsilon)$-approximation to \emph{clairvoyant}-CAP.
\end{proposition}

Similarly, one can show that when the number of segments $m$ is fixed, \emph{clairvoyant}-CAP has a ($1/2-\epsilon$)-approximation algorithm since \citet{el2021joint} proved the existence of a FPTAS for CAP in this case.

\subsection{Personalized and refined personalized assortments}\label{sec:p-taop}

Often a DCM is used to represent choices of heterogeneous consumer types as in  (\ref{eq:pm}). As discussed before, a p-TAOP-firm can identify the consumer types and personalize assortments is called a p-TAOP firm. Clearly a p-TAOP-firm can earn higher expected revenues than a TAOP-firm. The p-TAOP is also related to the personalized refined assortment optimization problem (p-RAOP) introduced by \citet{berbeglia2021refined}. Under the RAOP, a firm is allowed to make some products less attractive to avoid demand cannibalization. This is a more refined approach than simply removing such products as done in the TAOP. Likewise, a p-RAOP firm who can customize the refined assortment to each segment performs as least as well as the p-TAOP firm. However, not even the p-RAOP firm can do as well as the clairvoyant firm as it still has to deal with some residual uncertainty. We have shown in Corollary \ref{prop:personalized_by_segment} that $\bar{R} \leq 2 \cR^p$ under the LC-MNL model. Based on the above analysis and Corollary \ref{prop:personalized_by_segment}, we directly obtain the following result.


\begin{proposition}Under the LC-MNL model, 
$\cR^p \leq \cR^{p-raop} \leq \bar{\cR} \leq 2\cR^p$
where $\cR^{p-raop}$ denotes the optimal expected of a p-RAOP firm.
\end{proposition}

\citet{berbeglia2021refined} also provided a revenue guarantee for revenue-ordered assortments in relation to $\cR^{p-raop}$ in settings where each consumer type satisfies regularity. They showed that $\cR^{p-raop} \leq (1 + \ln(r_1/r_n))\cR^o$. Under any RUM, this bound also works with the clairvoyant firm (i.e. replacing $\cR^{p-raop}$ with $\bar{\cR}$) since we can interpret each joint realization of the product utilities as a different consumer type \footnote{There may be an infinite number of consumer types.} and each of them satisfies the regularity condition. Therefore

\begin{theorem}\label{RUM_prophet_prop} For every RUM,
$$\cR^p \leq \cR^{p-raop} \leq \bar{\cR} \leq [1 + \ln(r_1/r_n)]\cR^o.$$
\end{theorem}

Since it is possible to construct examples where $\cR^*$ can be made as close as possible to $(1+\ln(r_{max}/r_{min})) \cR^o$ (see \cite{berbeglia2020assortment}), the bound is tight.

Additionally, when we restrict to the LC-MNL model, we have that (i) $\cR^{raop} \leq \cR^p$ for every such model; and (ii) $\frac{\cR^p}{\cR^{taop}} \geq \min\{n,m\}$ for some LC-MNL models (see \citet{berbeglia2021refined}).

\ACKNOWLEDGMENT{The authors give thanks to Mina Irvani, Wentao Lu and Zhuodong Tang for their insightful comments. We are also very grateful to Danny Segev for giving us very valuable comments and pointing out relevant references. We are particularly grateful to Pin Gao for providing many detailed comments and suggestions to improve the paper's organization, exposition, and suggesting ideas to sharpen some results.}


\bibliographystyle{informs2014} 
\bibliography{references} 



\section{Appendix}

\textbf{Proof of Proposition~\ref{prop:mnl}} For any $S \subset N$ we have 
$$R(S) = R(S)\cdot \sum_{i \in S_+}\cM(i,S)  = \sum_{i \in S}r_i \cM(i,S).$$
Therefore,
$$R(S) \cM(0,S) = \sum_{i \in S}(r_i -R(S)) \cM(i,S).
$$
Dividing by $\cM(0,S)$ we obtain
$$R(S) = \sum_{i \in S}(r_i -R(S))v_i.$$

In particular, if $S^*$ maximizes $R(S)$ then $\cR^* = R(S^*)$ satisfies
$$\cR^* = \sum_{i \in S^*}(r_i -\cR^*)v_i.$$

It is easy to see that if there is an $i \in S^*$ such that $r_i - R(S^*) \leq 0$, then we can remove $i$ from $S^*$ and weakly improve the objective function so there is an optimal assortment $S^*$ such that $r_i > R(S^*)$ for all $i \in S^*$.  Moreover, if there is an $i \in N \setminus S^*$ with $r_i > R(S^*)$, then we can add $i$ to $S^*$ and strictly improve the objective function. Consequently there is an optimal assortment $S^*$ such that $S^* = \{i \in N: r_i > \cR^*\}$.

We now claim that $\cR^*$ is a root of the equation
$$\tau = \sum_{i \in N}(r_i -\tau)^+v_i.$$
This follows because 
\bean
\cR^* & = &  \sum_{i \in S^*}(r_i -\cR^*)v_i \\
 & = &  \sum_{i \in S^*}(r_i -\cR^*)v_i + \sum_{i \in N \setminus S^*}(r_i - \cR^*)^+v_i\\
& = & \sum_{i \in S^*}(r_i -\cR^*)^+v_i + \sum_{i \in N \setminus S^*}(r_i - \cR^*)^+v_i\\
& = & \sum_{i \in N}(r_i -\cR^*)^+v_i.
\eean
Since the left-hand term of $\tau = \sum_{i \in N}(r_i - \tau)^+ v_i$ is strictly increasing and the right hand term is decreasing it follows that the root is unique. 

By the implicit function theorem we see that 
$$\frac{\partial \cR^*_v}{\partial v_i} = \frac{(r_i - \cR^*_v)^+}{1 + \sum_{i \in N: r_i > \cR^*_v}v_i} \geq 0~~~~~\forall~~~~i \in N,$$
establishing that $\cR^*_v$ is weakly increasing in $v$. 
\Halmos

\textbf{Proof of Lemma~\ref{lem:formulaoddsratio}} 
For all $S \subset N$, $\cP(0,S) + \sum_{i \in S}\cP(i,S) = 1$. Consequently,
$$R(S) = \sum_{i \in S}r_i \cP(i,S) = R(S)[ \cP(0,S) + \sum_{i \in S}\cP(i,S)].$$
Subtracting $R(S)\sum_{i \in S}\cP(i,S)$ from the second term, and dividing by $\cP(0,S) \geq \cP(0,N) > 0$, yields the formula for $R(S)$. Applying the formula for $S^*$ yields the second result.\Halmos

\textbf{Proof of Proposition~\ref{prop:pi}}
The proof is based on the classical prophet inequality and given here for completeness.  The prophet inequality guarantees that there is a threshold, say $\tau$, such that the expected revenue of the decision maker who selects the first product with reward at least $\tau$, obtains a reward at least $0.5E[\max_{i \in N}X_i]$, when the $X_i$s are independent. The prophet inequality holds in particular for the case $X_i := r_iB_i$ with independent $B_i, i \in N$.

We remark that the above cited result result is independent of the order in which the decision maker considers the product, and holds in particular if the decision maker sees the products in the order $n, n-1, \cdots, 1$ and buys the first product, in this order whose reward exceeds the threshold.  By doing this the decision maker selects the lowest revenue product that exceeds the threshold. In contrast, if a non-clairvoyant firm offers the revenue-ordered assortment with threshold $\tau$, the consumer can select any of the products in the set $\{i \in N: r_i > \tau\}$ with $B_i = 1$, and may either select the same product as the decision maker or a more expensive one.

More formally, let $L(\tau)$ denote the expected revenue that the firm obtains when the consumer selects the first product with $r_i  > \tau$, that has $B_i = 1$, when the products are presented in the order $n, n-1, \cdots, 1$. This means that the decision maker obtains the smallest possible reward among all products in the revenue ordered assortment $S(\tau) : = \{i \in N: r_i > \tau\}$. From the classic prophet inequality we know that $L(\tau) \geq 0.5 \bar{\cR}$. To complete the proof we notice that $R(S(\tau))$, the expected reward from offering assortment $S(\tau)$, satisfies $R(S(\tau)) \geq L(\tau)$. This is because the consumer may select a higher revenue product, rather than the lowest revenue product in the set $S(\tau)$ with $B_i = 1$. Consequently, 
$$\cR^o \geq R(S(\tau)) \geq L(\tau) \geq 0.5E[\max_{i \in N}X_i] = 0.5 \bar{\cR}.$$

\Halmos



\textbf{Proof of Theorem~\ref{thm:newups}}
For any $\phi > 0$, we obtain an upper bound on $\bar{\cR}$ by recalling that $\bar{\cR} \leq \tau + \sum_{i \in N}(r_i - \tau)^+p_i$ and setting $\tau = \tau_{\phi}$,  the root of the equation $\tau = \sum_{i \in N}(r_i - \tau_{\phi})^+\phi p_i$.  This results in the bound
\bean
\bar{\cR} & \leq & \tau_{\phi} + \sum_{i \in N}(r_i - \tau_{\phi})^+p_i \\
& = &  \tau_{\phi} + \frac{1}{\phi}
\sum_{i \in N}(r_i - \tau_{\phi})^+\phi p_i\\
& = & \tau_{\phi} + \frac{1}{\phi}\tau_{\phi} = \frac{1 + \phi}{\phi}\tau_{\phi},
\eean
where the equality follows from the definition of $\tau_{\phi}$.
\Halmos

\textbf{Proof of Theorem~\ref{thm:sc}}

 Proof: Condition (\ref{eq:condphi}) is equivalent to  $\cP(i, S_{\phi}) \geq \phi p_i \cP(0,S_{\phi})$ for all $i \in S_{\phi}$.  Multiplying both sides by $(r_i - \tau_{\phi})$ and adding over  $i \in S_{\phi}$ we obtain
$$\sum_{i \in S_{\phi}}(r_i -\tau_{\phi}) \cP(i, S_{\phi}) \geq \cP(0,S_{\phi}) \sum_{i \in S_{\phi}} (r_i -\tau_{\phi}) \phi p_i = \cP(0,S_{\phi}) \tau_{\phi}$$
where the last equality follows since $\tau_{\phi}$ is the root of $\sum_{i \in N}(r_i - \tau)^+ \phi p_i = \tau$.
Moving the terms involving $\tau_p$ to the right we obtain
$$R(S_{\phi}) = \sum_{i \in S_{\phi}}r_i\cP(i, S_{\phi}) \geq \tau_{\phi}\cdot \left[\cP(0, S_{\phi}) + \sum_{i \in S_{\phi}} \cP(i, S_{\phi})\right] = \tau_{\phi}.$$
Combining this with the upper-bound in Theorem~\ref{thm:newups} we obtain:
$$\tau_{\phi} \leq R(S_{\phi}) \leq \cR^* \leq \bar{\cR} \leq \frac{1 + \phi}{\phi}\tau_p,$$
so the prophet inequality holds for $\phi$, and also for $\phi^*$. Moreover, since $S_{\phi}$ is a revenue-ordered assortment we have $R(S_{\phi}) \leq \cR^o \leq \cR^*$, so revenue-order heuristic also satisfies $\cR^o \geq \frac{\phi^*}{1 + \phi^*} \bar{\cR}$.
 \Halmos



If (\ref{eq:condphi}) holds for $\phi$, then following the same logic we obtain the set of inequalities
$$\cR^*_{\phi \omega} \leq R(S_{\phi \omega}) \leq \cR^* \leq \bar{\cR} \leq 2 \cR^*_{\omega}.$$

Notice that for $\phi \in (0,1)$,
$$\phi \cR^*_{\omega} = \sum_{i \in S^*_{\omega}} \frac{r_i \phi \omega_i}{1 + \sum_{i \in S^*_{\omega}} \omega_i}
\leq \sum_{i \in S^*_{\omega}} \frac{r_i \phi \omega_i}{1 + \sum_{i \in S^*_{\omega}} \phi \omega_i} \leq \cR^*_{\phi \omega},$$
implying that $\cR^*_{\omega} \leq \cR^*_{\phi \omega}/\phi$. 
Consequently
$$\cR^*_{\phi \omega} \leq  \cR^* \leq \bar{\cR} \leq 2 \cR^*_{\omega} \leq \frac{2}{\phi} \cR^*_{\phi \omega},$$
so
$$\frac{\bar{\cR}}{\cR^*} \leq \frac{2}{\phi}$$
as claimed. Since $\cR^*_{\phi \omega} \leq \cR^o \leq \cR^*$ it also follows that $\cR^o \leq  \bar{\cR} \leq  2\cR^o/\phi$, 
so
$$\frac{\bar{\cR}}{\cR^o} \leq \frac{2}{\phi}$$
as claimed. \Halmos

\textbf{Proof of Theorem~\ref{thm:fcr}}  We will first show that under Assumption R, $\mbox{Pr}(B_i = 0, i \in S) = \cP(0,S)$.  We will do this by showing that the events $B_i = 0, i \in S$ and the event that $0$ is selected from $S$ are equivalent. 

Let $C = \{i \in N: B_i = 1\}$. Suppose that $B_i = 0$ for all $i \in S$  but $0$ is not selected from $S$. Then there is a product $i \in S$ that is elected when $S$ is offered, so $\hat{B}_i(S)  = 1$. By R1,  $\hat{B}_i(S) =1$ implies that $B_i = 1$ so $i \in C \cap S$ which is a contradiction. Conversely, assume that $0$ is selected from $S$, but $C \cap S$ is non-empty.  Then by R2, $0$ is not selected from $S$ which is a contradiction. Since the two events are equivalent we have $\mbox{Pr}(B_i = 0, i \in S) = \cP(0,S)$.

We now apply this result to assortments $[i-1]$ and $[i]$. Notice that $\mbox{Pr}(B_j = 0, \forall j \in [i-1])  = \mbox{Pr}(B_j = 0, \forall j \in [i-1], B_i = 0) +  \mbox{Pr}(B_j = 0, \forall j \in [i-1], B_i =1)$. If $C$ is non-empty, we define $m(B): = \min\{i: i \in C\}$, otherwise we set $m(B) : = 0$. Clearly $\max_{i \in N}X_i = X_{m(B)}$, so a key to computing $\bar{\cR}$ is to find the distribution of $m(B)$. Clearly,
\bean
\mbox{Pr}(m(B) = i) & = &  \mbox{Pr}(B_j = 0, \forall j < i, B_i =1)\\
& = &  \mbox{Pr}(B_j = 0, \forall j \in [i-1]) - \mbox{Pr}(B_j = 0, \forall  j \in [i])\\
& = & \cP(0,[i-1]) - \cP(0,[i]),
\eean
so
\bean
\bar{\cR}  & =  & \sum_{i \in N} r_i\cdot \cP(m(B) = i)\\
& = &  \sum_{i \in N} r_i\cdot \left[\cP(0,[i-1]) - \cP(0,[i])\right].
\eean
where the last equality follows from simple algebra by setting $r_{n+1} = 0$. 
 \Halmos




\textbf{Proof of Theorem~\ref{thm:ubn}} We assume that the revenues are decreasing in $i$.  By R1, if a firm offers the best revenue-ordered assortment, it earns 

 \bean
\cR^o  & = & \max_{i \in N} \sum_{j \in [i]}r_j \cP(j,[i])\\
& \geq  & \max_{i \in N} \sum_{j \in [i]}r_i \cP(j,[i])\\
& = & \max_{i \in N} r_i \sum_{j \in [i]}\cP(j,[i])\\
& = & \max_{i \in N} r_i(1 - \cP(0,[i]))\\
& \geq & \max_{i \in N} r_i(1 - \cP(0,\{i\}))\\
& = & \max_{i \in N} r_i\omega_i,
\eean
where the first inequality follows because the $r_i$ are decreasing in $i$, and the second from R1 which implies that $\cP(0,\{i\}) \geq \cP(0,[i])$.
On the other hand, by R1 and R2 we have
\bean
\bar{\cR} & = & E[r_{m(B)}]\\
& = & \sum_{i \in N}r_i \mbox{Pr}(m(B) = i)\\
& \leq & \sum_{i \in N}r_i \mbox{Pr}(B_i = 1)\\
& = & \sum_{i \in N}r_i \omega_i \\
& \leq & n \max_{i \in N}r_i \omega_i,
\eean
where the first inequality follows since for all $i \in N$, the event $m(B) = i$ implies the event $B_i = 1$ and therefore $\mbox{Pr}(m(B) = 1) \leq \mbox{Pr}(B_i = 1)$. The second equality follows by R1, and the last  is straightforward. This establishes that
$$\bar{\cR} ~\leq ~n \max_{i \in N} r_i \omega_i ~\leq ~ n \cR^o ~\leq ~ n \cR^*$$

We now provide an example where  
$$\cR^p = \bar{\cR} ~\geq ~n \cR^* - \epsilon$$
for any $\epsilon > 0$. Suppose there are $m = n+1$ consumer types, and all the types have deterministic utilities for products $i \in N$ that satisfy $u_1 < u_2 < \cdots < u_n$. The consumer types are differentiated by the utility they assign to the  outside alternative, or alternatively by their preference ordering when the outside alternative is included. More precisely, for type $j = i$, product $i$ is the smallest index product that ranks higher than the outside alternative. Thus consumers of type 1 have preference ordering $0 \prec 1 \prec \cdots \prec n$,  type two consumers have preference ordering $1 \prec 0 \prec 2 \cdots \prec n$, and so on, with type $n$ consumers having preference ordering $1 \prec \cdots \prec n-1 \prec 0 \prec n$.  Type $0$ consumers are those with preference ordering $1 \prec \cdots \prec n-1  \prec n \prec 0$.  Let $\theta_j$ be the proportion of consumers of type $j \in M$, with $\sum_{j \in M} \theta_j = 1$. 

Consider the revenue-ordered assortment $[i]$. While product $i$ is the highest utility product in $[i]$, only consumers of types $j \in [i]$ will prefer $i$ over the outside alternative. Thus $R([i]) = r_i \sum_{j \leq i} \theta_j$ for all $i \in N$. Clearly $\omega_i = \cP(i, \{i\}) = \sum_{j \leq i} \theta_j$, so $R([i]) = r_i \omega_i, i \in N$, and 
$\cR^o = \max_{i \in N} r_i \omega_i$. We will next show that $\cR^* = \cR^o$. To see this, notice that $R(S) = r_{i(S)}\omega_{i(S)}$ where $i(S) = \max\{i \in S\}$. Therefore $\max_{S \subset N}R(S) = \max_{i \in N}r_i \omega_i = \cR^o$. 

We  now turn to the computation of $\bar{\cR}$ and $\cR^p$. For this purpose notice that $\cP(0,[i]) = 1 - \omega_i$, so from Theorem~\ref{thm:fcr} we have 
$$\bar{\cR} = \sum_{i \in N}r_i( \omega_i -  \omega_{i-1}) = 
\sum_{i \in N}r_i \theta_i.$$
Consider now the p-TAOP firm that can identify the type of each arriving consumer. Such a firm will offer product $i$ to type $i$ consumers earning 
$$\cR^p = \sum_{i \in N}r_i \theta_i = \sum_{i \in N}r_i( \omega_i -  \omega_{i-1})= \bar{\cR}.$$

We will now show how to set the parameters $r_i, \omega_i, i \in N$ so that the bound holds and is tight in the limit.  Set $r_i =   \omega_1/ \omega_i, i \in N$. Since $\omega_i = \sum_{j \leq i}\theta_j$ is increasing in $i$, we see that $r_i$ is decreasing in $i \in N$. Since $r_i \omega_i =  \omega_1$ for all $i \in N$ it follows that $\cR^o = \cR^* =  \omega_1$ while
\bean
\bar{\cR}  & =  & \sum_{i \in N}r_i( \omega_i -  \omega_{i-1})\\
& = &  \sum_{i \in N}\frac{ \omega_1}{ \omega_i}( \omega_i -  \omega_{i-1})\\
& = & \sum_{i \in N} \omega_1 \left(1 - \frac{ \omega_{i-1}}{ \omega_i}\right).
\eean
Set $ \omega_i := \delta^n(\delta^{-i}-1)$. Then  $ \omega_i$ is increasing in $i$ with $ \omega_n = 1 - \delta^n < 1$. Then

\bean
\frac{\bar{\cR}}{\cR^*}   & =  & \sum_{i \in N}\left(1 - \frac{ \omega_{i-1}}{ \omega_i}\right)\\
& = & \sum_{i \in N}\frac{1 - \delta}{1- \delta^i}
\eean
with the sum converging to $n$ as $\delta \downarrow 0$, showing that for sufficiently small $\delta$ we have $\bar{\cR} \geq n \cR^* - \epsilon$. 

Next, we show that the RUM model constructed above can be represented by an instance of the Markov chain model (although every Markov chain can be represented as a RUM \citep{berbeglia_markov}, the converse is not always the case). To see this, consider a Markov chain model with $\lambda_i = 0, i \in [n-1]$ and $\lambda_n = \omega_n$. Suppose further that the only possible transitions from $i$ are to either $i-1$ or the outside alternative $0$. Consequently, for $i = 1$ we have $\rho_{1,0} = 1$. For $i > 1$, set  $\rho_{i,i-1}  = \omega_{i-1}/\omega_i$, and $\rho_{i,0} = 1 - \omega_{i-1}/\omega_i$. Then clearly, for any $S \subset N$, $P(i,S) = 0$ if $i \neq i(S)$ and for $i = i(S)$ we have
$$\mathcal{P}(i,S) = \omega_n \prod_{j=i+1}^{n}\frac{\omega_{j-1}}{\omega_j} = \omega_i$$
 Then $\cR^* = \max_{i \in N}r_i \omega_i$. On the other hand, for $S = [i]$ we have $\cP(i,[i]) = \omega_i$, so $\cP(0,[i]) = 1 - \omega_i$, and by Theorem~\ref{thm:fcr}, we have  $\bar{\cR} = \sum_{i \in N} r_i(\omega_i -  \omega_{i-1})$. Then selecting the $\omega_i = \delta^n(\delta^{-i}-1)$, we see that the ration converges to $n$ even for the MC model.  \Halmos

\textbf{Proof of Proposition~\ref{prop:idm}} Notice that  $R(S) = \sum_{i \in S}r_i \lambda_i$ is increasing in $S$, so 
$$\cR^* = \max_{S \subset N}R(S) = R(N) = \sum_{i \in N}r_i \lambda_i,$$
implying that $\cR^* = \cR^p = \bar{\cR}$.
\Halmos


\textbf{Proof of Theorem~\ref{thm:mnl}}
From the upper bound of Theorem~\ref{thm:newups} applied to an MNL model with attraction vector $v$ and $v_0 = 1$, we know that
 $$\cR^o = \cR^*_v \leq \bar{\cR}_v \leq 2 \cR^*_{\omega},$$
where for the MNL
 $$\omega_i = \frac{v_i}{1 + v_i} \leq v_i~~~\forall~~i \in N.$$
From Proposition~\ref{prop:mnl} we know that $\cR^*_v$ is increasing in $v$, so $\omega \leq v$ implies that
 $\cR^*_{\omega} \leq \cR^*_v = \cR^o$. Consequently 
 $$\bar{\cR}_v \leq 2 \cR^*_{\omega} \leq 2 \cR^*_v = 2\cR^o.$$
To see that the bound is tight consider an MNL with  $n = 2$, $r_1 = 1$ and $r_2  =r_1\cM(1, \{1\}) = v_1/(1+v_1)$. Then, $S^*= \{1\}$ and $\cR^*=  r_2$ while
$$\bar{\cR}_v = [1 + \cM(0,S_1)\cM(2,S_2)]\cR^*.$$
Consequently,
$$\frac{\bar{\cR}_v}{\cR^*_v}  = 1 + \frac{1}{1+v_1}\frac{v_2}{1+v_1+v_2} \rightarrow 2$$
as $v_1 \downarrow 0$ and $v_2 \rightarrow \infty$.

\Halmos

\textbf{Proof of Theorem~\ref{thm:bigbeta}}

For the MNL model with $v = e$, $O(i,S) = 1$ for all $i \in S, S \subset N$. On the other hand, $\omega_i = 0.5$ for all $i \in N$. Therefore, the inequality $O(i,S) \geq \phi \omega_i$ holds for $\phi = 2$ for all $i \in S, S \subset N$. By Theorem~\ref{thm:sc} we have 
$$\cR^*_e \geq \frac{\phi}{1 + \phi} \bar{\cR} = \frac{2}{3} \bar{\cR}$$
from which we see that $\bar{\cR} \leq 1.5 \cR^*_e$.




\textbf{Proof of Theorem~\ref{thm:GAM}} 
Notice that the odds-ratios for the GAM are given by

$$O(i,S) = \frac{v_i}{1 + \sum_{j \in N \setminus S}w_j}$$
which is increasing in $S$. This implies that $O(i,S) \geq O(i, \{i\}) \geq \omega_i$ for all $S \ni i$, so by Theorem~\ref{thm:sc} the prophet inequality (\ref{eq:piphi}) holds for the GAM with $\phi = 1$. Since the MNL is a special case of the GAM it follows that the GAM is also tight. 
\Halmos

\textbf{Proof of Corollary~\ref{prop:personalized_by_segment}}
Offering personalized assortment $S^*_{v_j}$ for class $j \in M$ yields an expected revenue of
$$\sum_{j \in M}\theta_j \cR^*_{v_j} \leq \sum_{j \in M}\theta_j \bar{\cR}_{v_j} \leq
2 \sum_{j \in M}\theta_j \cR^*_{v_j} $$
where the second inequality follows from Theorem~\ref{thm:mnl}. \Halmos

\textbf{Proof of Proposition~\ref{prop:dcm}}. Let $g^*$ be the smallest vector such that $g \geq r$ and $g \geq \rho g$, where $\rho$ is the transition matrix of the MC. It has been shown, see \citep{blanchet2016markov}, that an optimal assortment is given by $S^* = \{i \in N: g_i = r_i\}$. If $i \in S^*$ then a firm prefers to issue product $i$ and earn $r_i$ rather than foregoing $i$ and continuing in the MC until absorption in $S^*_+$. On the other hand, if $i \notin S^*$ then the firm prefers to $r_i$, by not offering it, and then allow the consumer to take as many Markov steps as needed until absorption  in $S^*_+$. 

The optimal expected revenue is shown to be equal to
$$\sum_{i \in N}\lambda_i g^*_i = \sum_{i \in S^*}\lambda_i r_i + \sum_{i \notin S^*}\lambda_i g^*_i.$$
Notice that in this formula $g^*_i$ is the maximum revenue that we can obtain from consumers whose first-choice is $i$, and this is achieved by offering the non-personalized assortment $S^*$. As a result, no further benefits can be obtained by personalizing based on first-choices.
\Halmos

\textbf{Proof of Lemma~\ref{lem:nlm}}:

Suppose that our first claim is false. Then for all $k \in L(S)$, we have $[R_k(S_k) - R(S)]V_k(S_k) < 0$, so it must be that $V(S_k) > 0$, and therefore that $R_k(S_k) < R(S)$.  Let $T$ be constructed as suggested in the statement of the lemma. Then
$R(T)$ is the unique root of the equation 
$$\tau = \sum_{k \notin L(S)}[R_k(S_k) - \tau]\frac{V_k(S_k)}{v_0}$$
We next show that at $\tau = R(S)$ we have
$$R(S) <  \sum_{k \notin L(S)}[R_k(S_k) - R(S)]\frac{V_k(S_k)}{v_0}$$
which implies that the root $R(T) > R(S)$. To see this notice that 
\bean
  R(S)  & =  &  \sum_{k \in L(S)}[R_k(S_k) - R(S)]\frac{V_k(S_k)}{v_0} + \sum_{k \notin L(S)}[R_k(S_k) - R(S)]\frac{V_k(S_k)}{v_0}\\
  & < & \sum_{k \in L(S)}[R_k(T_k) - R(S)]\frac{V_k(T_k)}{v_0} + \sum_{k \notin L(S)}[R_k(S_k) - R(S)]\frac{V_k(S_k)}{v_0}
\eean
which establishes the first part of the lemma.  Suppose now that $L(S) = \emptyset$. Then 
 $$R(S) = \sum_{k \in K}[R_k(S_k) - R(S)]V_k(S_k)/v_0 = 
 \sum_{k \in K}[R_k(S_k) - R(S)]_+V_k(S_k)/v_0,$$
 so $R(S)$ is the unique root of the claimed equation establishing the second part of the lemma. 
  \Halmos

\textbf{Proof of Theorem~\ref{thm:nlm}}: From Lemma~\ref{lem:nlm} we know that if  
 $$L(S^*) = \{k \in K: [R_k(S^*_k) - \cR^*]V_k(S^*_k) < 0\}$$
is non-empty, then we can find a $T$ such that $R(T) > R(S^*) = \cR^*$, but this would contradict the optimality of $S^*$, and therefore it must be that $L(S^*)$ is empty. We have established that for all $k \in K$, $[R_k(S^*_k) - \cR^*]V_k(S^*_k) \geq 0$. This implies that either $R_k(S^*_k) \geq R(S^*)$ or $S^*_k = \emptyset$. Therefore 
$$R(S^*)  =  \sum_{k \in K}[R_k(S^*_k) - R(S^*)] V_k(S^*_k)/v_0 = \sum_{k \in K}[R_k(S^*_k) - R(S^*)]_+ V_k(S^*_k)/v_0$$
is unique root of the equation
 $$\tau = \sum_{k \in K}[R_k(S^*_k) - \tau]_+ V_k(S^*_k)/v_0.$$
 \Halmos


\textbf{Proof of Theorem~\ref{thm:nlsc}}: 

We have already established that 
$$\cR^* ~=~  \sum_{k \in K}R_k(S^*_k)Q(k,S^*) ~\leq~ \bar{\cR}.$$

Suppose that $\bar{\cR} = \bar{R}(\bar{S})$ for some assortment $\bar{S}$.
We now provide the desired upper bound for $\bar{\cR}$.
\bean
\bar{\cR} & = & E[\max_{k \in K}~X_k(\bar{S}_k)]\\
& \leq &  \cR^* + \sum_{k \in K}[R_k(\bar{S_k})- \cR^*]_+ Q(k, \bar{S}_k)\\
& \leq &  \cR^* + \sum_{k \in K}[R_k(\bar{S}_k) - \cR^*]_+ \frac{V_k(\bar{S}_k)}{v_0}\\
& \leq & 2\cR^*,
\eean
where the first inequality follows since $E[\max_{k \in K}X_k(\bar{S}_k)] \leq \tau + \sum_{k \in K}E[X_k(\bar{S}_k) -\tau]_+$ for all $\tau$ and selecting $\tau = \cR^*$. The second inequality follows from $Q(k, \bar{S}_k) \leq V_k(\bar{S}_k)/v_0$ for all $k$. 

To establish the last inequality consider the set $L(\bar{S})$. If the set is empty then we set $\bar{T} = \bar{S}$; otherwise we construct $\bar{T}$ from $\bar{S}$ by setting $\bar{T}_k = \emptyset$ for all $k \in L(\bar{S})$. Clearly $R(\bar{T}) \geq R(\bar{S})$ by Lemma~\ref{lem:nlm}. From this construction we see that 
$$\sum_{k \in K}[R_k(\bar{S}_k) - \cR^*]_+ \frac{V_k(\bar{S}_k)}{v_0} = 
\sum_{k \in K}[R_k(\bar{T}_k) - \cR^*]_+ \frac{V_k(\bar{T}_k)}{v_0}$$
on account of $\cR^* \geq R(\bar{T})$. Suppose for a contradiction that 
$$\sum_{k \in K}[R_k(\bar{T}_k) - \cR^*]_+ \frac{V_k(\bar{T}_k)}{v_0} > \cR^*.$$ Then we can construct an assortment based on $\bar{T}$, that sets $\bar{T}_k = \emptyset$ for all $k$ such that $R_k(\bar{T}_k) < \cR^*$, with expected revenue higher than $\cR^*$ which is a contradiction. Therefore, it must be that 
$$\sum_{k \in K}[R_k(\bar{T}_k) - \cR^*]_+ \frac{V_k(\bar{T}_k)}{v_0} \leq  \cR^*$$
as claimed. \Halmos


\textbf{Proof of Proposition~\ref{prop:pricing_unbounded}}
Suppose there is a single product with gross utility $U$ and that $U_0 = 1$ almost surely. Suppose further that
$$\cP(U \geq p) = \min\{1,1/p\}~~~~0 < p \leq 1+a$$
for some constant $a > 0$, with $\cP(U \geq p) = 0$ for $p > 1 + a$. The clairvoyant firm observes $U$ and prices at  $p = U$ obtaining expected profit $\bar{\cR} = E[U] = (1+\ln(1+a))$. On the other hand, the non-clairvoyant firm obtains profit $\cR^* = 1$ by using any price $p \in [1, 1+a]$. The ratio is therefore $1 + \ln(1+a)$ which can be made arbitrarily large as $a \rightarrow \infty$.\Halmos

\textbf{Proof of Proposition~\ref{prop:MNLprice}}
From our analysis for the general case we know that
$$\bar{\cR} = E[(U_N-U_0)^+] = E[\max(U_N, U_0)] - E[U_0] = E[\max(U_N, U_0)]$$
when $E[U_0] = 0$ as in the normalized MNL case. Thus, for the MNL
$$\bar{\cR} = E[\max(U_N, U_0)] = \ln(1 + V(N)),$$
where $V(N) = \sum_{i \in N}e^{u_i}$ \citep{gumbel1935valeurs}.
For the non-clairvoyant firm, it is well known that $p_i = p$ for all $i \in N$, so
 $$\cR^* = \max_p \frac{pV(N)}{\exp(p) + V(N)}.$$

It is easy to show that at optimality $p$ satisfies that $(p-1)\exp(p) = V(N)$, so that $\cR^* = p-1$. Since the optimal profit is positive, it must be that an optimal price is at least 1.
Thus,

$$\frac{\bar{\cR}}{\cR^*} = \frac{\ln(1 + (p-1)\exp(p))}{p-1} = \frac{\ln(1 + x\exp(x+1))}{x} = f(x),$$
where $x \in (0,\infty)$.

The derivative is $$f'(x)= \frac{\frac{\exp(x+1)x(x+1)}{\exp(x+1)x+1} - \ln(\exp(x+1)x + 1)}{x^2},$$
and one can verify that $f'(x) < 0$ for all $x>0$ so $f(x)$ is a decreasing function.
Thus,

$$\frac{\bar{\cR}}{\cR^*} \leq \lim_{x \to 0^+}\frac{\ln(1 + x\exp(x+1))}{x} = e =\exp(1).$$

We can also see that the bound is tight in the limit when $p \downarrow 1$, which occurs when $V(N) \downarrow 0$. \Halmos

\textbf{Proof of Proposition~\ref{prop:factor_2:joint}}
Let $T^*_{c-CAP}$, $T^*_{CAP}$ denote the optimal assortments chosen in the first stage for the \emph{clairvoyant}-CAP and CAP respectively.
\begin{eqnarray*}
\mathcal{R}^*_{clairvoyant-cap} & = & \sum_{j \in M}\theta_j \bar{\cR}_j(T^*_{c-CAP})\\
& \leq & \sum_{j \in M}\theta_j \max_{S \subseteq T^*_{c-CAP}} 2 \cdot \mathcal{R}_j(S)\\
& \leq & \sum_{j \in M}\theta_j \max_{S \subseteq T^*_{CAP}}2 \cdot \mathcal{R}_j(S)\\
& = & 2 \cdot \mathcal{R}^*_{cap}
\end{eqnarray*}
The first inequality follows from Theorem~\ref{thm:mnl} and the second from the optimality of assortment $T^*_{CAP}$. \Halmos

\textbf{Proof of Proposition~\ref{clairvoyant_CAP_np-hard}}
Observe that in the case where all products have the same prices, the clairvoyant expected revenue ($\mathcal{R}^*_{clairvoyant-cap}$), is the same as the CAP revenue ($\mathcal{R}^*_{cap}$). Since \citet{el2021joint} proved that CAP is NP-hard even in the case where all revenues are the same, the result follows. \Halmos

\end{document}